\documentclass[preprint,12pt,authoryear]{elsarticle}
\usepackage{graphicx}
\usepackage{soul}

\graphicspath{{./figures/}, {./figures/supplementary/}}

\usepackage{gensymb} 
\usepackage{hyperref}
\usepackage{multirow}
\usepackage{bm}
\usepackage{amsmath,amssymb}
\usepackage{cleveref}
\usepackage{soul}
\usepackage{float}
\usepackage[labelformat=simple]{subcaption}
\DeclareCaptionLabelFormat{bold}{\textbf{#2} }

\usepackage[usenames,dvipsnames]{color}

\newcommand{\norm}[1]{\left\lVert#1\right\rVert}

\def\equationautorefname{Eq.}

\def\myfigautorefname~#1\null{Fig.~S#1\null}

\def\equationautorefname~#1\null{%
  Eq.~(#1)\null
}

\usepackage{algpseudocode}
\usepackage{algorithm}

\algrenewcommand\algorithmicindent{0.7em}

\algnewcommand\algorithmicswitch{\textbf{switch}}
\algnewcommand\algorithmiccase{\textbf{case}}
\algdef{SE}[SWITCH]{Switch}{EndSwitch}[1]{\algorithmicswitch\ #1\ \algorithmicdo}{\algorithmicend\ \algorithmicswitch}%
\algdef{SE}[CASE]{Case}{EndCase}[1]{\algorithmiccase\ #1}{\algorithmicend\ \algorithmiccase}%
\algtext*{EndSwitch}%
\algtext*{EndCase}%
\algnewcommand{\LineComment}[1]{\State \(\triangleright\) #1}
\makeatletter
\renewcommand{\Function}[2]{%
  \csname ALG@cmd@\ALG@L @Function\endcsname{#1}{#2}%
  \def\@currentfunction{#1}%
}
\newcommand{\funclabel}[1]{%
  \@bsphack
  \protected@write\@auxout{}{%
    \string\newlabel{#1}{{\@currentfunction}{\thepage}}%
  }%
  \@esphack
}
\makeatother

\newif\ifcomment
\commenttrue

\newcommand{\figsize}{0.4}

\newcommand{\Grad}{\nabla}
\newcommand{\GradAdim}{\nabla'}
\newcommand{\Div}{\nabla \cdot}
\newcommand{\DivAdim}{\nabla' \cdot}
\newcommand{\Laplacian}{\nabla^{2}}
\newcommand{\LaplacianAdim}{\nabla^{\prime 2}}

\newcommand{\Space}{x}
\newcommand{\Spacex}{x}

\newcommand{\Spacez}{z}
\newcommand{\bmSpace}{\bm{\Space}}
\newcommand{\bmSpaceAdim}{\bm{\Space}'}
\newcommand{\Lx}{L_x}
\newcommand{\Lz}{L_z}
\newcommand{\LengthScale}{\Lz}
\newcommand{\LxLz}{A}

\newcommand{\Time}{t}
\newcommand{\TimeScale}{\tau}
\newcommand{\TimeAdim}{\Time'}

\newcommand{\di}[1]{\mathop{\mathrm{d}#1}}

\newcommand{\Normal}{\bm{n}}

\newcommand{\SpaceTime}{\left( \bmSpace, \Time \right)}

\newcommand{\GravityCt}{g}
\newcommand{\GravityVec}{\bm{g}}
\newcommand{\GravityUnit}{\bm{e}_g}
\newcommand{\Angle}{\theta}

\newcommand{\UnitVecx}{\bm{e}_x}
\newcommand{\UnitVecz}{\bm{e}_z}

\newcommand{\Density}{\rho}
\newcommand{\DensityRef}{\Density_0}
\newcommand{\DensityScale}{\tilde{\Density}}
\newcommand{\DensityAdim}{\Density'}
\newcommand{\DensityConc}{\Density \left( \Concentration \right)}

\newcommand{\Porosity}{\phi}

\newcommand{\DarcyVelocSymbol}{q}
\newcommand{\DarcyVelocScale}{\tilde{\DarcyVelocSymbol}}
\newcommand{\DarcyVeloc}{\bm{\DarcyVelocSymbol}}
\newcommand{\DarcyVelocAdim}{\DarcyVeloc'}

\newcommand{\Source}{s}

\newcommand{\PermSymbol}{k}
\newcommand{\PermTensor}{\bm{\PermSymbol}}
\newcommand{\PermScale}{\tilde{\PermSymbol}}
\newcommand{\PermGeoMean}{\PermSymbol_g}
\newcommand{\PermTensorAdim}{\PermTensor'}

\newcommand{\Viscosity}{\mu}
\newcommand{\ViscosityAdim}{\Viscosity'}
\newcommand{\ViscosityScale}{\tilde{\Viscosity}}

\newcommand{\Pressure}{p}

\newcommand{\PressureAdim}{\Pressure'}
\newcommand{\PressureRef}{\Pressure_0}

\newcommand{\diffVol}{\di \Omega}
\newcommand{\diffSurf}{\di \Gamma}

\newcommand{\Domain}{\Omega}

\newcommand{\Boundary}{\partial \Domain}

\newcommand{\Concentration}{c}
\newcommand{\ConcentrationNeutral}{\Concentration_n}
\newcommand{\ConcentrationMaximum}{\Concentration_m}
\newcommand{\VarianceConcentration}{\sigma^2_{\Concentration}}

\newcommand{\Diffusion}{D}
\newcommand{\Dispersion}{\bm{\Diffusion}}
\newcommand{\Rayleigh}{\text{Ra}}

\newcommand{\COO}{\ensuremath{\mathrm{CO_2}}}

\newcommand{\MaxDensity}{\Density_{m}}

\newcommand{\DeltaDensitym}{\Delta \MaxDensity}

\newcommand{\Mass}{M}
\newcommand{\DomainBuoyant}{\Domain_b}
\newcommand{\MassBuoyant}{M_b}
\newcommand{\InterfaceLength}{L_I}

\newcommand{\ScalarDissipationRate}{\chi}

\newcommand{\Mixing}{f} 

\newcommand{\Curve}{\mathcal{C}}
\newcommand{\CurveTime}{\Curve_{\Time}}
\newcommand{\Tangent}{\bm{t}}
\newcommand{\NormalLine}{\gamma}

\newcommand{\MomentLine}[1]{m_{#1}}

\newcommand{\MeanLogK}{\mu_{\log \PermSymbol}}

\newcommand{\MySigma}{\sigma}
\newcommand{\Variance}{\MySigma^2_{\log \PermSymbol}}
\newcommand{\CorrLenx}{\lambda_{\Spacex}}
\newcommand{\CorrLenz}{\lambda_{\Spacez}}
\newcommand{\RatioCorrLen}{\gamma}
\newcommand{\VarianceLine}{\MySigma^2}

\newcommand{\Width}{w}
\newcommand{\WidthAverage}{\langle \Width \rangle}

\newcommand{\OF}{OpenFOAM}
\newcommand{\SF}{SECUReFOAM}

\newcommand{\NElemz}{N_z}
\newcommand{\NElemx}{N_x}
\newcommand{\ElemSize}{h}

\usepackage{xr} 
\usepackage{tikz}
\usetikzlibrary{positioning}
\usetikzlibrary{calc}
\usepackage[mathlines]{lineno}
\usepackage{xcolor}

\journal{Advances in Water Resources}
\begin{document}
\begin{frontmatter}
\title{Mixing and spreading of gravity currents in heterogeneous porous media}%
\author[authorETSIAE,authorIDAEA]{Albert Jiménez-Ramos}\ead{albert.jimenez.ramos@upm.es}
\author[authorIDAEA]{Juan J. Hidalgo}\ead{juanj.hidalgo@csic.es}
\cortext[cor1]{Corresponding author}
\address[authorETSIAE]{ETSIAE-UPM-School of Aeronautics, Universidad Politécnica de Madrid, Madrid, Spain}
\address[authorIDAEA]{Institute for Environmental Assessment and Water
  Research, Spanish National Research Council, Barcelona, Spain}

%
%
\begin{abstract}
  We analyze the mixing, migration and spreading of a gravity current in a heterogeneous porous medium using high-fidelity numerical simulations. Heterogeneity is represented by log-normal permeability fields of varying correlation lengths and variance. Stable and unstable density stratification scenarios are considered through linear and non-monotonic density laws, respectively. In the stable case, heterogeneity increases dissolution at early times but the late-time behavior is similar to the homogeneous scenario. In the unstable case, heterogeneity accelerates the onset of convection. Convection-driven dissolution slows the gravity current and counteracts the dispersive effect of heterogeneity resulting in a narrower interface and higher dissolution than in the stable case. These effects are monotonic with respect to the Rayleigh number. Permeability anisotropy reduces dissolution because of the barrier effect of low permeability regions, except when blobs of buoyant fluid are trapped in low permeability structures and rapidly dissolve. A larger variance of the log-permeability field enhances dissolution. However, the homogeneous case outperforms heterogeneous cases except when the Rayleigh number is small. This suggests an interaction between the size of the instabilities, the correlation structure of the permeability field, and the dispersive and barrier effects of the permeability field that controls dissolution efficiency.
\end{abstract}
%
%
%
%
\begin{highlights}
\item Dispersive effect of heterogeneity is counteracted by the fingering instability
\item Anisotropy reduces dissolution due to barrier effect except for trapped blobs
\item Dissolution enhanced only for high variance permeability fields
\end{highlights}
%
%
%
%
\end{frontmatter}
%
%
%
%
\section{Introduction}
\label{sec:Intro}

Gravity currents form when the flow is driven by density differences between fluids. Sea water intrusion in which the dense salt water encroaches on the fresh water that discharges into the sea \citep{Cooper1959}, aquifer remediation of DNAPLs using dense brines \citep{Miller2000}, or \COO{} injection in saline aquifers, in which the supercritical \COO{} migrates over the resident brine \citep{bachuAquiferDisposalCO21994}, are examples of the formation of gravity currents in porous media.

Gravity currents are often studied under the assumption that fluids are immiscible and the medium is homogeneous. In that case, the speed and shape of the gravity current are determined by the density and viscosity contrast between fluids and aquifer permeability \citep{Huppert1995, Gasda2011}. However, there are scenarios in which these assumptions can be too strict. Immiscibility neglects the formation of a mixing zone between fluids as, for example, the one formed by salt water and fresh water in coastal aquifers \cite{Abarca2009}, while natural porous media are always heterogeneous \citep{Tian2024}. 

The effects of fluid miscibility and medium heterogeneity have been independently addressed in previous works. On one had, it has been shown that mixing and dissolution of the fluids change the amount of migrating fluid, reducing the size and extension of the gravity current \citep{macminnSpreadingConvectiveDissolution2012, szulczewskiEvolutionMiscibleGravity2013}. This is especially important when the density stratification is unstable, as in \COO{} injection, and the formation of fingering instabilities enhances fluid dissolution, which can eventually arrest the migration of the gravity current \citep{hidalgoDynamicsConvectiveDissolution2013, MacMinn2013buoyant, eleniusInteractionsGravityCurrents2015}. On the other hand, the effect of permeability heterogeneity on gravity currents also changes their shape and spread. In layered media, gravity currents tend to focus on high permeability layers \citep{leahyApplicationGravityCurrents2009, Huppert2013}, increasing their speed \citep{benhamTwophaseGravityCurrents2021}, and the migration of the current depends on the cross-bedded structure of the formation \citep{Whelan2023}. In media with vertical permeability gradients, gravity currents can have a rarefaction wave or shock-like structure
\citep{hintonBuoyancydrivenFlowConfined2018}. In general, vertical and horizontal permeability gradients alter the gravity current's spread, average height and steepness \citep{cirielloEffectVariablePermeability2013}, while homogeneous anisotropic permeability can delay the formation of the current \citep{Benham2022}.

A few works studied the combined effect of miscibility and medium heterogeneity on gravity currents. Heterogeneity influences how fluids mix by widening the size of the mixing zone between the fluids \citep{AbarcaTesis, Pool2015} and favoring dispersion \citep{Kerrou2009, Flicos2025}. Therefore, altering the migration and shape of the gravity current. The further effect of heterogeneity on convective instabilities can greatly modify the migration and evolution of the gravity current. However, it has only been studied for horizontal stratification of fluids \citep{Ranganathan2012, Green2018, mahyapourEffectPermeabilityHeterogeneity2022, Guo2023-3D, benhammadiExperimentalNumericalStudy2025, benhammadiEffectPermeabilityHeterogeneity2025a, Guo2025}.

We focus here on the combined effect of heterogeneity and dissolution on gravity currents without capillarity effects. Neglecting capillarity is the strongest assumption of this work \citep[see][for a detailed discussion of capillarity]{hidalgoDynamicsConvectiveDissolution2013} as it neglects the formation of a capillary fringe, residual trapping of the fluid, and capillary pressure hysteresis. However, it helps us isolate the effect of heterogeneity and convective instabilities on transport processes.

The goal of this work is to analyze the impact of the porous medium heterogeneity on the migration and dissolution of a gravity current. To that end, we analyze the shape of the fluid interface, and the evolution of dissolution fluxes by means of high-resolution numerical simulations. We consider two scenarios. A stable gravity current in which density changes linearly with concentration and an unstable gravity current in which a non-monotonic density stratification triggers fingering instabilities.

The work is organized as follows. \autoref{sec:governingEquations} presents the governing equations \autoref{sec:methodology} describes the numerical setup and defines the metrics to quantify mixing and spreading. In \autoref{sec:results}, we report and discuss the obtained results. Finally, some concluding remarks are provided in \autoref{sec:conclusions}.
%
%
\section{Governing equations}
\label{sec:governingEquations}

We discretize a semi-infinite aquifer using a two-dimensional rectangular domain $\Domain$ in the $x$-$z$ plane, with dimensions $\Domain = \left[ 0, \LxLz \right] \times \left[ 0, 1 \right]$. In the following, we describe the governing equations in dimensionless form and refer to \autoref{sec:dimensionlessProcedure} for the nondimensionalization procedure \citep{riazOnsetConvectionGravitationally2006}.

We consider two perfectly miscible fluids and neglect capillarity effects so that fluid mass conservation, assuming incompressible fluid flow, under the Boussinesq approximation, and in the absence of sources or sinks, reads 
\begin{equation}
	\Div \DarcyVeloc = 0,
	\label{eq:incompressibility}
\end{equation}
where
\begin{equation}
	\DarcyVeloc = - \PermTensor \left( \Grad \Pressure + \Density \UnitVecz \right)
	\label{eq:Darcy}
\end{equation}    
is the Darcy velocity, $\PermTensor \equiv \PermSymbol \left( \bmSpace \right) \bm{I}$ is the permeability tensor, assumed isotropic, where $\PermSymbol \left( \bmSpace \right)$ is a scalar function and $\bm{I}$ is the identity tensor, $\Pressure \equiv \Pressure \SpaceTime$ is the pressure, $\UnitVecz$ is a unit vector parallel to gravity and pointing in the opposite direction, and $\Density$ is the fluid density, which is a function of concentration.

Mass transport is described by the advection-diffusion equation in terms of a scaled concentration $\Concentration$, representing the ratio between fluids. Thus, the transport equation in a Eulerian framework in dimensionless form reads
\begin{equation}
\frac{\partial \Concentration}{\partial \Time} = - \DarcyVeloc \cdot \Grad \Concentration + \frac{1}{\Rayleigh} \Laplacian \Concentration,
\label{eq:transport}
\end{equation}
where the Rayleigh number $\Rayleigh$ is defined as
\begin{equation}
  \label{eq:Ra}
  \Rayleigh = \frac{\PermGeoMean \Delta \Density \GravityCt \Lz}{\Porosity \Diffusion \Viscosity},
\end{equation}
with $\PermGeoMean$ the geometric mean of the permeability field, $\Delta \Density$ a characteristic density difference, $\GravityCt$ the gravity constant, $\Lz$ the vertical extension of the domain, $\Porosity$ the porosity, $\Viscosity$ the fluid viscosity, considered constant, and $\Diffusion$ the dispersion coefficient. The dispersion coefficient accounts for molecular diffusion and hydrodynamic dispersion and we assume that it is a scalar and constant (see \autoref{sec:dimensionlessProcedure}).

Regarding boundary conditions, pressure is prescribed along the right boundary and the other boundaries are taken to be impervious,
\begin{subequations}
    \label{eq:boundaryConditions}
    \begin{align}
	\Pressure = 0 &\text{ at } \Spacex = \LxLz, \\
	\DarcyVeloc \cdot \Normal = 0 &\text{ at } \Boundary \setminus \lbrace \Spacex = \LxLz \rbrace, \\
	\Grad \Concentration \cdot \Normal = 0 &\text{ at } \Boundary,
\end{align}
\end{subequations}
where $\Normal$ denotes the outward normal vector in $\Boundary$.

Initially, the region $x\leq \frac{\LxLz}{5}$ is filled with the buoyant fluid ($\Concentration=1$). We choose to fill the region $x \leq \frac{\LxLz}{5}$ to model the system configuration after the injection phase and avoid the boundary condition at $x=\LxLz$ to affect the current (see Fig. S1 in the supplementary material). During the simulated time the gravity current's nose position stays below $x=12$ for all cases. That is, the gravity current is confined to the left half of the domain. This can be seen in the figures displaying the nose position (Fig. S2--S6 in the supplementary material) and velocity magnitude and pressure profiles along the $x$-axis for the least favorable case (homogeneous, $\Rayleigh=10000$, Figure S1 in the supplementary material). To render the initial sharp interface smooth and ease numerical convergence, we set
\begin{equation}
        \Concentration \left( \bmSpace, 0 \right) = \frac{1}{2} \left( 1 - \tanh \left( 100 \left( \Spacex - \frac{\LxLz}{5} \right) \right) \right) \quad \text{in } \Domain, 
        \label{eq:initialCondition_concentration}
\end{equation}
where $\tanh(x)$ is used as a differentiable approximation to the Heaviside step function describing the initial interface (see \autoref{sec:numericsDetails}).
We assume that, after injection, the fluids are at rest, $\DarcyVeloc \left( \bmSpace, 0 \right) = \bm{0}$, and we use $\Pressure \left( \bmSpace, 0 \right) = 0$ as an initial guess for the pressure.

%
%
\subsection{Constitutive laws}


We consider two different constitutive laws for the fluid density: one leading to a stable density configuration and one leading to an unstable density configuration. The \emph{stable case} has a linear density law $\DensityConc = -3.6\Concentration$ that represents a stable fluid stratification, similar to the behavior of salt water and fresh water (with $\Concentration$ the salt content). The slope $-3.6$ is chosen so that the buoyant force is the same as in the unstable configuration. The \emph{unstable case} has a density law that aims to capture the relevant aspects of \COO{}-brine systems. That is, a density law with a density contrast that favors the fluids stratification and drives the migration of the gravity current, and with an intermediate density maximum that triggers fingering instabilities and drives convective dissolution. We use a dimensionless density law based on an analog fluid system formed by propylene glycol and water \citep{backhausConvectiveInstabilityMass2011, hidalgoDynamicsConvectiveDissolution2013}
\begin{equation}
	\DensityConc = 6.19 \Concentration^3 - 17.86 \Concentration^2  + 8.07 \Concentration.
	\label{eq:cubicDensityLaw}
\end{equation}

\begin{figure}[t]
	\centering
	\includegraphics[width=0.5\textwidth]{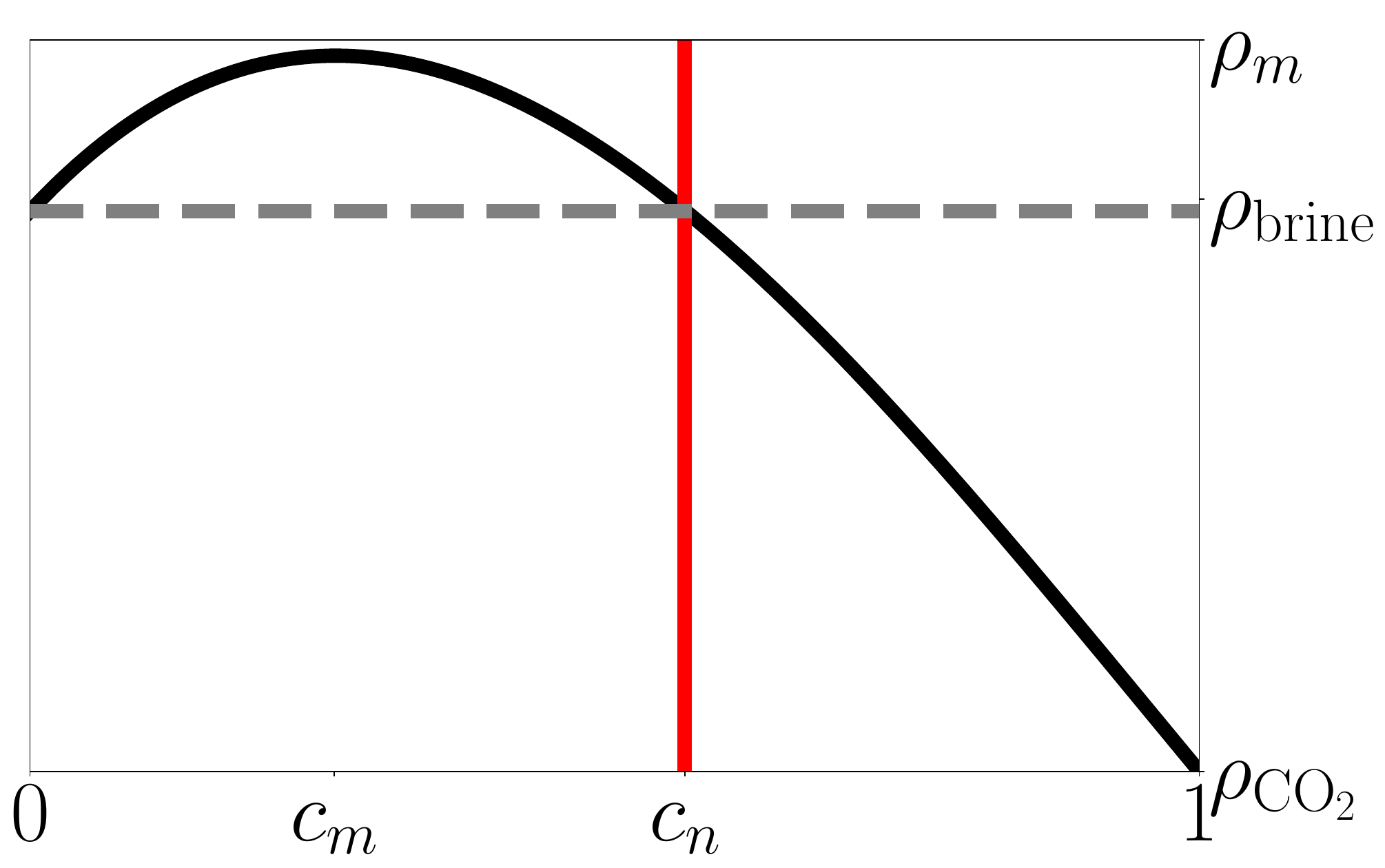}
	\caption{Graph of the constitutive density law $\Density \left( \Concentration \right)$ in \autoref{eq:cubicDensityLaw} with a neutral concentration at $\ConcentrationNeutral = 0.56$. When $\Concentration > \ConcentrationNeutral$, the fluid tends to float, while when $\Concentration < \ConcentrationNeutral$, the fluid  tends to sink.}
	\label{fig:cubicDensityLaw}
\end{figure}
This density law (see \autoref{fig:cubicDensityLaw}) is a non-monotonic function of the concentration, actually the mixing fraction for this case, with an intermediate maximum at $\ConcentrationMaximum = 0.26$. There exists a neutral concentration $\ConcentrationNeutral = 0.56$ such that the density of the mixture is equal to the density of the ambient fluid. This contour acts as an interface because when the fluid has concentration $\Concentration > \ConcentrationNeutral$, it is less dense than the ambient fluid and tends to float, but when $\Concentration < \ConcentrationNeutral$, it is denser than the ambient fluid and tends to sink. We use $\Concentration = \ConcentrationNeutral$ to separate buoyant mobile fluid from the dense mixture that sinks.
%
%
\section{Methodology}
\label{sec:methodology}
%
%
\subsection{Heterogeneous permeability fields}
\label{sec:generationPermeability}

We consider a two-dimensional heterogeneous porous medium represented by a spatially-varying permeability field $\PermSymbol \left( \bmSpace \right)$. The permeability is modeled as a log-normal stochastic process characterized by its mean $\MeanLogK$, variance $\Variance$, and correlation lengths $\CorrLenx, \CorrLenz$ in the $\Spacex$ and $\Spacez$ coordinates, respectively. The spatial correlation is quantified using an exponential autocorrelation function \citep{gelharThreedimensionalStochasticAnalysis1983}.

\begin{figure}[t]
  \centering
  \begin{subfigure}[b]{\textwidth}
    \centering
    \caption{}
    \label{fig:permeabilitites_iso}
    \includegraphics[width=\textwidth]{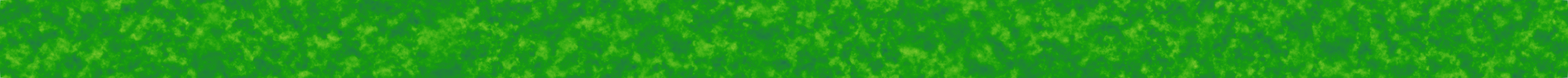}
  \end{subfigure}
  \begin{subfigure}[b]{\textwidth}
    \centering
    \caption{}
    \label{fig:permeabilitites_aniso2_var1}
    \includegraphics[width=\textwidth]{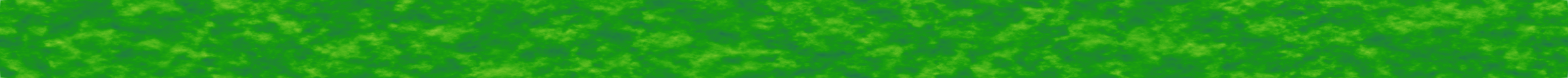}
  \end{subfigure}
  \begin{subfigure}[b]{\textwidth}
    \centering
    \caption{}
    \label{fig:permeabilitites_aniso2_var5}
    \includegraphics[width=\textwidth]{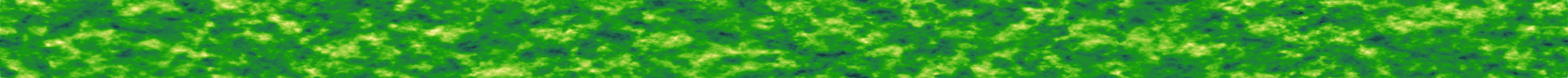}
	
	\end{subfigure}
	\centering
        \includegraphics[width=0.4\textwidth]{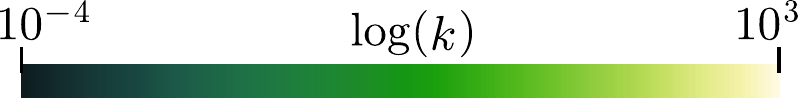}
	\caption{Permeability field $\PermSymbol \left( \bmSpace
          \right)$ obtained as the realization of a random field
          following a log-normal distribution with zero mean, and
          \subref{fig:permeabilitites_iso} isotropic correlation
          lengths $\CorrLenx=\CorrLenz=0.05$ and variance
          $\Variance=1$, \subref{fig:permeabilitites_aniso2_var1}
          anisotropic correlations lengths $\CorrLenx=0.1$ and
          $\CorrLenz=0.05$ and variance $\Variance=1$, and
          \subref{fig:permeabilitites_aniso2_var5} anisotropic
          correlations lengths $\CorrLenx=0.1$ and $\CorrLenz=0.05$
          and variance $\Variance=5$. Note that the fields \subref{fig:permeabilitites_aniso2_var1} and \subref{fig:permeabilitites_aniso2_var5} correspond to the same realization but with different variance $\Variance$.}
	\label{fig:permeabilityField}
      \end{figure}

To generate the desired permeability field, we use the spectral algorithm of \cite{ruanEfficientMultivariateRandom1998}. We first generate a Gaussian random field with zero mean and unit variance and the desired spatial correlation structure $\CorrLenx$ and $\CorrLenz$, and then, we apply an exponential transformation to obtain a log-normal random field whose logarithm has zero mean and the desired variance $\Variance$. This method is convenient because it allows to reuse the realization for different $\Variance$ while keeping the same spatial structure. In \autoref{fig:permeabilityField}, we show two realizations of the permeability field with different values of $\CorrLenx$, $\CorrLenz$ and $\Variance$. Note that \autoref{fig:permeabilitites_aniso2_var1} and \autoref{fig:permeabilitites_aniso2_var5} correspond to the same realization with different $\Variance$.
%
%
\subsection{Numerical simulations setup}
\label{sec:numericalSimulationsSetup}
We solve equations (\ref{eq:incompressibility} -- \ref{eq:transport}) using the finite volume method. Specifically, we use \SF{} package \citep{icardiComputationalFrameworkComplex2023} implemented in \OF{} \citep{wellerTensorialApproachComputational1998}. Time derivatives are discretized using the second-order implicit backward scheme with adaptive time-stepping. For the divergence terms we use a second-order van Leer limited scheme to ensure that the concentration values remain bounded between 0 and 1, while for the Laplacian terms we use the second-order scheme. 

\newcommand{\RefValueMesh}{\alpha}
The domain $\Domain = \left[ 0, \LxLz \right] \times \left[ 0, 1 \right]$, with $\LxLz = 20$, is discretized using a structured quadrilateral mesh composed of $\NElemz=800$ cells along the vertical direction and $\NElemx = \LxLz \NElemz = 16000$ cells along the horizontal direction, which gives an element size $\ElemSize = 0.00125$. To determine the needed mesh resolution, we perform a mesh-convergence study and compare the results with those obtained using a finer mesh (see details in \autoref{sec:numericsDetails}). The comparison shows that, in the worst case, the relative error with respect to the reference value is approximately below $10^{-1}$. Using a mesh with element size $\ElemSize$, the simulations are run until the non-dimensional time $\Time = 20$ and take between 2 and 14 days to run in a cluster with 100 CPUs. The data for a single case occupies approximately 150 GB. Thus, we find our discretization to be a good compromise between accuracy and computational memory and time.
%
%
\subsection{Metrics for characterizing spreading, migration and mixing}
\label{sec:metrics}
We analyze the time evolution of the buoyant current with six macroscopic quantities: the mass of buoyant fluid, the length and width of the fluid interface, the total dissolution rate, the average dissolution flux per unit length of the current interface and the scalar dissipation rate.

The scalar dissipation rate is defined as
\begin{equation}
\ScalarDissipationRate \left( \Time \right) = \frac{1}{\Rayleigh} \int_{\Domain} \norm{\Grad \Concentration \SpaceTime}^2 \di{\Omega}.
\label{eq:scalarDissRate}
\end{equation}
Since the simulated system can be considered semi-infinite because the gravity current remains far from the right boundary during the simulated time interval (see the pressure and velocity magnitude near the boundary in Fig. S1 in the supplementary material), the same relation between the variance of concentration $\VarianceConcentration$ and the scalar dissipation holds, that is,  $\frac{\partial \VarianceConcentration}{\partial \Time} = - 2 \ScalarDissipationRate$ \citep{popeTurbulentFlows2000,jhaFluidMixingViscous2011}. Thus, $\ScalarDissipationRate$ decreases as concentration homogenizes and indicates the rate at which fluids mix.

Since there is no actual interface separating the buoyant and ambient fluid because we are considering two miscible fluids, we use the properties of the non-monotonic density law \autoref{eq:cubicDensityLaw} to define a separating region. The fluid mixture tends to float when $\Concentration > \ConcentrationNeutral$ and tends to sink when $\Concentration < \ConcentrationNeutral$ (see \autoref{fig:cubicDensityLaw}). Therefore, we define the amount of buoyant mass relative to the injected mass as
\[
\MassBuoyant \left(\Time \right) = \frac{1}{\Mass} \int_{\DomainBuoyant \left( \Time \right)} \Concentration \SpaceTime \diffVol,
\]
where $\Mass$ is the initial mass of buoyant fluid and $\DomainBuoyant \left(\Time \right) = \lbrace \bmSpace \in \Domain \vert \Concentration \SpaceTime > \ConcentrationNeutral \rbrace$. When the density law is linear, we set $\ConcentrationNeutral = 0.5$. This value corresponds to the maximum mixing between the fluids.

The interface separating the buoyant fluid from the dense fluid mixture is calculated using the contour $\Concentration = \ConcentrationNeutral$. In practice, the length of the poly-line that approximates this contour determines the interface length, denoted by $\InterfaceLength \left( \Time \right)$.

The total dissolution rate is given by $-\frac{\di{\MassBuoyant}}{\di{\Time}}$. To have an average measure of the dissolution rate along the interface, we compute the average dissolution flux per unit length as $-\frac{1}{\InterfaceLength}\frac{\di{\MassBuoyant}}{\di{\Time}}$. In practice, we compute $\frac{\di{\MassBuoyant}}{\di{\Time}}$ by finite differences and apply a Savitzky-Golay filter \citep{savitzkySmoothingDifferentiationData1964,jonesSciPyOpenSource2001} to smooth the noise originated in numerical precision of the solver.

\renewcommand{\figsize}{0.95}
\begin{figure}[t]
	\centering
	\begin{subfigure}[b]{\figsize\textwidth}
	\begin{tikzpicture}
	\node[inner sep=0pt, outer sep=0pt] (img) {\includegraphics[width=\textwidth]{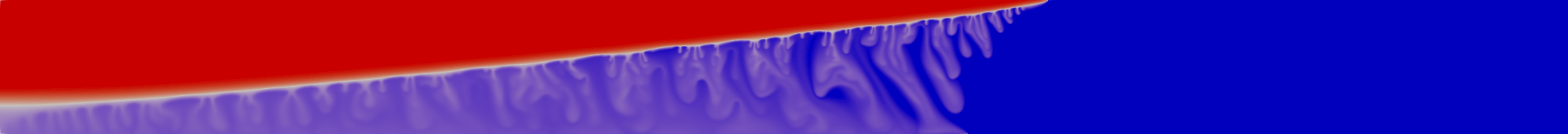}};
	\path let
	\p1 = ($(img.north)-(img.south)$)
	in
	node[right=0pt of img]
	{\includegraphics[height=\y1]{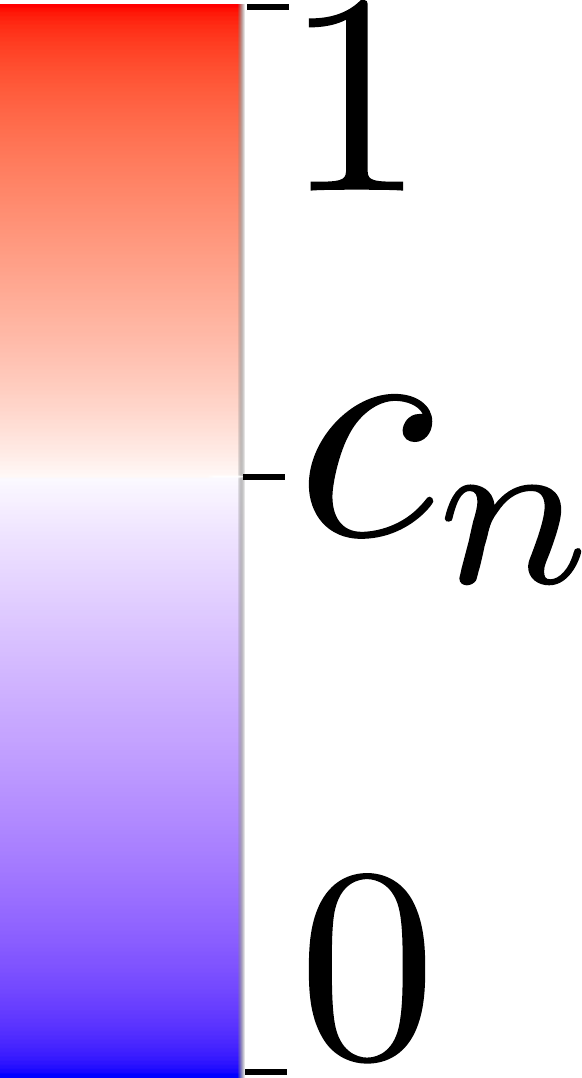}};
\end{tikzpicture}
	\caption{}
	\label{fig:homo_Ra5000_70}
	\end{subfigure} \\
	\begin{subfigure}[b]{\figsize\textwidth}
	\begin{tikzpicture}
		\node[inner sep=0pt, outer sep=0pt] (img) {\includegraphics[width=\textwidth]{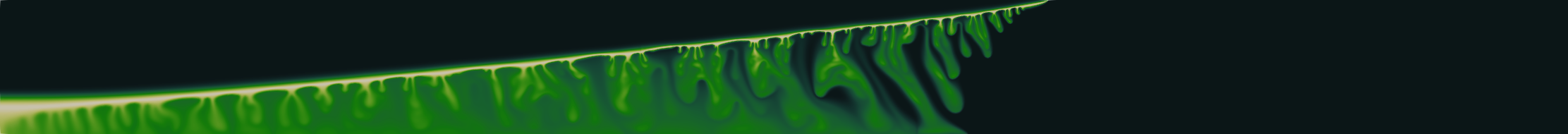}};
		\path let
		\p1 = ($(img.north)-(img.south)$)
		in
		node[right=0pt of img]
		{\includegraphics[height=\y1]{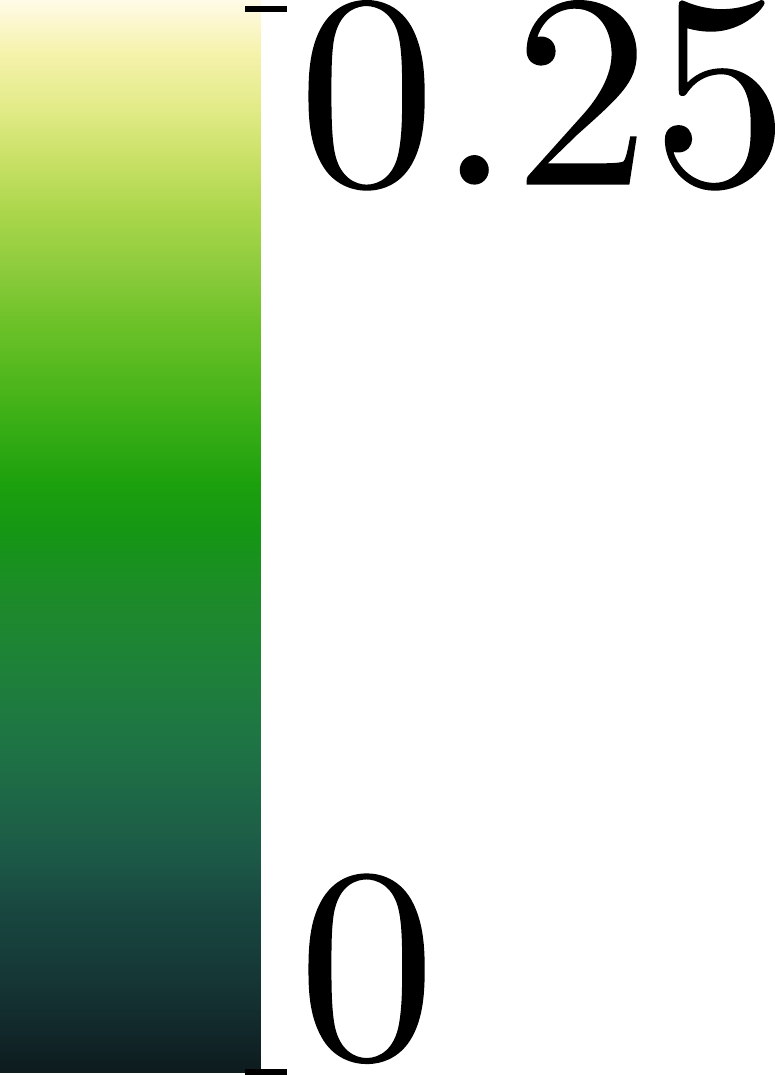}};
	\end{tikzpicture}
	\caption{}
	\label{fig:homo_Ra5000_mixing_70}
\end{subfigure}
	\caption{Snapshot of the simulation of a gravity current in a homogeneous porous medium at $\Time = 7$. \subref{fig:homo_Ra5000_70} Concentration field $\Concentration \SpaceTime$. \subref{fig:homo_Ra5000_mixing_70} Field $\Mixing \SpaceTime = \Concentration \SpaceTime \left( 1 - \Concentration \SpaceTime \right)$. The domain extends to $\Spacex = 20$, but only $0 \leq \Spacex \leq 12$ is shown here.}
\label{fig:c_and_c1c}
\end{figure}

To determine the interface width, we approximate the mixing region by $\Mixing\SpaceTime = \Concentration \SpaceTime \left[ 1 - \Concentration \SpaceTime \right] $ (see \autoref{fig:c_and_c1c}). The function $\Mixing \SpaceTime$ decays to zero away from the mixing zone and attains its maximum at $\Concentration = 0.5$. When the mixing region evolves perpendicular to the horizontal axis, the second centered moment of $\Mixing$ along lines parallel to the $\Spacex$-axis gives a good estimation of the interface width \citep{perezAssessmentPredictionPoreScale2020}. However, when gravitational effects exist, the interface rotates and this method fails.

\renewcommand{\figsize}{0.45}
\begin{figure}[t]
    \centering
    \includegraphics[width=\linewidth]{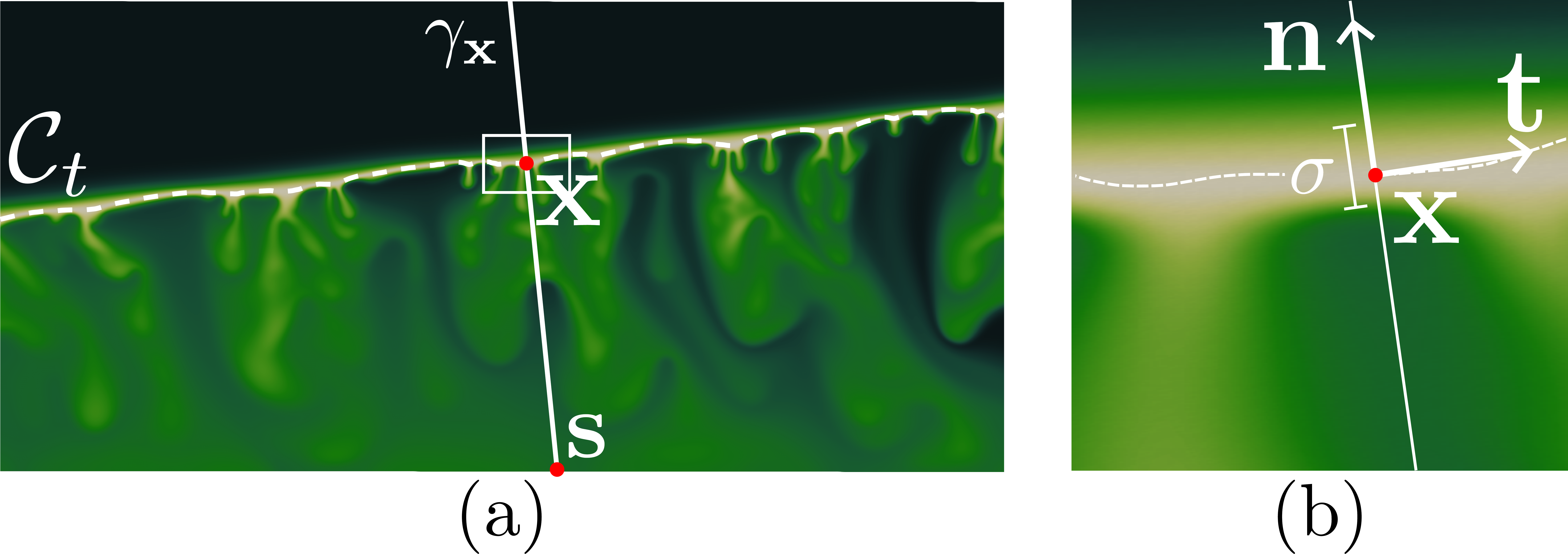}
    {\phantomsubcaption\label{fig:mixingScheme_contour}}
    {\phantomsubcaption\label{fig:mixingScheme_close}}
   	\caption{Mixing interface width. \subref{fig:mixingScheme_contour} Zoom-in of the interface showing the curve $\CurveTime$ in a dashed white line, the line $\NormalLine_{\bmSpace}$ at a point $\bmSpace \in \CurveTime$, and the intersection point $\bm{s}$ between $\NormalLine_{\bmSpace}$ and $\Boundary$. \subref{fig:mixingScheme_close} A closer view illustrating the tangent $\Tangent_{\CurveTime, \bmSpace}$ and normal $\Normal_{\CurveTime, \bmSpace}$ vectors. The value $\MySigma \left( \Time ; \bmSpace \right)$ estimates the interface width.}
    \label{fig:mixingScheme}
\end{figure}

To compute the width of a rotating interface, we consider the curve 
\begin{equation}
	\CurveTime = \lbrace \bmSpace \in \Domain \vert \Concentration \SpaceTime = 0.5 \rbrace
	\label{eq:curveContour}
\end{equation}
and  denote by $\Tangent_{\CurveTime, \bmSpace}$ and $\Normal_{\CurveTime, \bmSpace}$ the unit tangent and normal vectors to $\CurveTime$ at a point $\bmSpace \in \CurveTime$ (see \autoref{fig:mixingScheme}). The orientation of $\Tangent_{\CurveTime, \bmSpace}$ is chosen such that $\Tangent_{\CurveTime, \bmSpace} \cdot \UnitVecx \geq 0$. Next, we define the normal line to the curve $\CurveTime$ at $\bmSpace$ as
\begin{equation}
	\NormalLine_{\bmSpace} = \lbrace \bm{y} \in \Domain \vert \left( \bm{y} - \bmSpace \right) \cdot \Tangent_{\CurveTime, \bmSpace} = 0 \rbrace
	\label{eq:normalLine}
\end{equation}
(see \autoref{fig:mixingScheme_contour}). Then, we define the $i$-th moment of $\Mixing$ along the line $\NormalLine_{\bmSpace}$ as
\[
\MomentLine{i} \left( \Time ; \bmSpace \right) = \int_{\NormalLine_{\bmSpace}} \left[ \left(\bm{y} - \bm{s} \right) \cdot \Normal_{\CurveTime, \bmSpace} \right]^i \Mixing \left( \bm{y}, \Time \right) \di{\bm{y}},
\]
where $\bm{s}$ is arbitrarily one of the two intersection points between $\NormalLine_{\bmSpace}$ and $\Boundary$ (see \autoref{fig:mixingScheme_contour}). Thus, the interface width at a point $\bmSpace \in \CurveTime$ along the line $\NormalLine_{\bmSpace}$ is given by the second centered moment
\[
\VarianceLine \left( \Time ; \bmSpace \right) = \frac{\MomentLine{2} \left( \Time ; \bmSpace \right)}{\MomentLine{0} \left( \Time ; \bmSpace \right)} - \left( \frac{\MomentLine{1} \left( \Time ; \bmSpace \right)}{\MomentLine{0} \left( \Time ; \bmSpace \right)} \right)^2
\]
(see \autoref{fig:mixingScheme_close}). Finally, the effective interface width along the whole interface is defined by averaging over the curve $\CurveTime$,
\begin{equation}
	\WidthAverage \left( \Time \right) = \frac{\int_{\CurveTime} \MySigma \left( \Time ; \bmSpace \right) \diffSurf}{\int_{\CurveTime} \diffSurf}.
	\label{eq:effectiveWidthContour}
\end{equation}

When fingers develop or the interface is distorted due to the heterogeneity, integrating along the line $\NormalLine_{\bmSpace}$ may overestimate the interface width since $\NormalLine_{\bmSpace}$ may intersect the curve $\CurveTime$ several times. To avoid this issue, in practice we filter the points on $\NormalLine_{\bmSpace}$ and only consider those up to the first minimum of $\Mixing$ on each side of $\bmSpace$. For very tortuous interfaces though, this heuristic may fail and the width may be overestimated.

%
%
\section{Results}
\label{sec:results}

In this section, we consider the problem described in \autoref{sec:governingEquations} for the two density laws and different properties of the permeability field. First, we study the migration and mixing of a gravity current for the stable and unstable cases in homogeneous porous media, \autoref{sec:new_homogeneous}, and in isotropic heterogeneous media, \autoref{sec:new_isotropic}. Next, in \autoref{sec:anisotropy}, we analyze the effect of different correlation-length anisotropy values $\CorrLenx$ and $\CorrLenz$. Finally, in \autoref{sec:variance}, we study the impact of different variances $\Variance$.

When we consider heterogeneous media, the reported results correspond to the average of two realizations. Statistically converged trends require an average over hundred of realizations, however, this is unaffordable given the computational burden of the simulated scenarios (see \autoref{sec:numericalSimulationsSetup}). To overcome this shortcoming we considered permeability fields with at least 20 correlation lengths in any direction, so that the gravity current and fingering instability can sample the permeability structure.
%
%

\subsection{Gravity currents in homogeneous porous media}
\label{sec:new_homogeneous}

\renewcommand{\figsize}{0.4}
\begin{figure}[t]
    \centering
    \includegraphics[width=\textwidth]{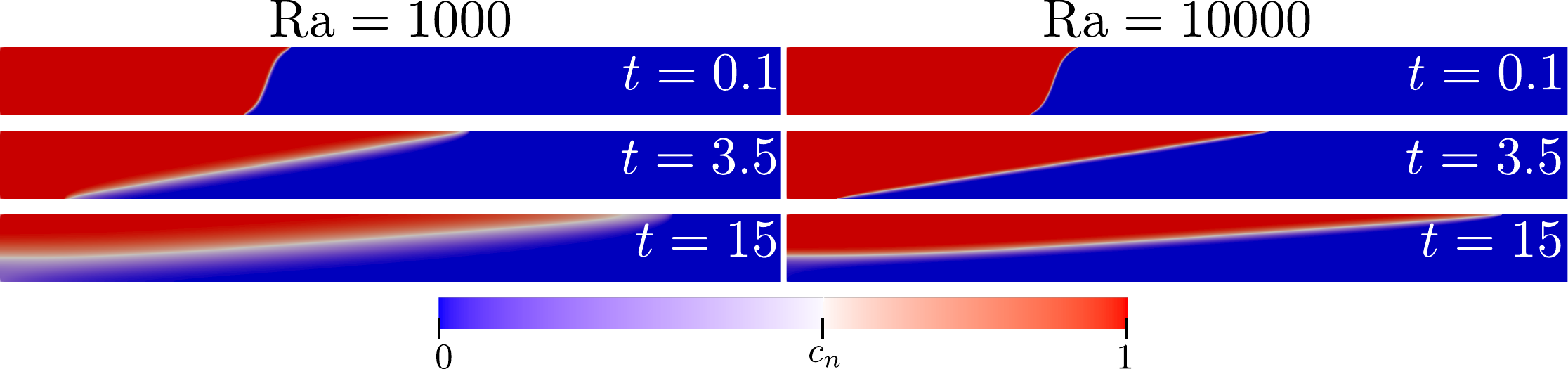}
    \caption{Snapshots of the concentration field for the stable case in a homogeneous medium for $\Rayleigh = 1000$ (left) and $\Rayleigh = 10000$ (right) at times $\Time = 0.1$ (top), $\Time = 3.5$ (middle), and $\Time = 15$ (bottom). The domain extends to $\Spacex = 20$, but only $0 \leq \Spacex \leq 12$ is shown here.}
    \label{fig:comparisonGravityCurrentHomoStable}
\end{figure}

\renewcommand{\figsize}{0.4}
\begin{figure}[t]
    \centering
    \includegraphics[width=\textwidth]{homo_unstable_Ra1K_vs_10K}
    \caption{Snapshots of the concentration field for the unstable case in a homogeneous medium for $\Rayleigh = 1000$ (left) and $\Rayleigh = 10000$ (right) at times $\Time = 0.1$ (top), $\Time = 3.5$ (middle), and $\Time = 15$ (bottom). The domain extends to $\Spacex = 20$, but only $0 \leq \Spacex \leq 12$ is shown here.}
    \label{fig:comparisonGravityCurrentHomoUnstable}
\end{figure}

In this section, we study the migration and mixing of a gravity current in a homogeneous porous medium. At initial times, the stable and unstable cases are qualitatively similar, but differences appear when instabilities occur (see Movie M1 in the supplementary material).

First, lateral diffusion dominates and the interface transitions from vertical to an s-shape due to buoyancy forces (see \autoref{fig:comparisonGravityCurrentHomoStable} and \autoref{fig:comparisonGravityCurrentHomoUnstable} for $\Time = 0.1$). Next, as the interface rotates, it transitions from an s-shape into a tilted straight line. In the stable case, the interface widens as the fluids start to mix due to diffusion (see \autoref{fig:comparisonGravityCurrentHomoStable} for $\Time = 3.5$ and $\Time = 15$). In the unstable case, the mixing of both fluids leads to the formation of a dense fluid layer, which triggers the fingering instability and enhances dissolution (see \autoref{fig:comparisonGravityCurrentHomoUnstable} for $\Time = 3.5$). Finally, the detachment of the plume from the bottom happens earlier in the unstable case than in the stable case (see Movie M1 in the supplementary material). In the former case, the layer of dense fluid accumulated at the bottom suppresses convective dissolution along a progressively larger fraction of the interface \citep{hidalgoDynamicsConvectiveDissolution2013} (see \autoref{fig:comparisonGravityCurrentHomoUnstable} for $\Time = 15$). For low $\Rayleigh$, the interface is wider and the high-density mixture fingers are thicker and get dissolved before reaching the bottom of the domain. In contrast, for high $\Rayleigh$, the interface is sharper and the fingers become thinner and longer (see \autoref{fig:comparisonGravityCurrentHomoUnstable}). 

\renewcommand{\figsize}{0.32}
\begin{figure}[t]
        \centering
        \includegraphics[width=\linewidth]{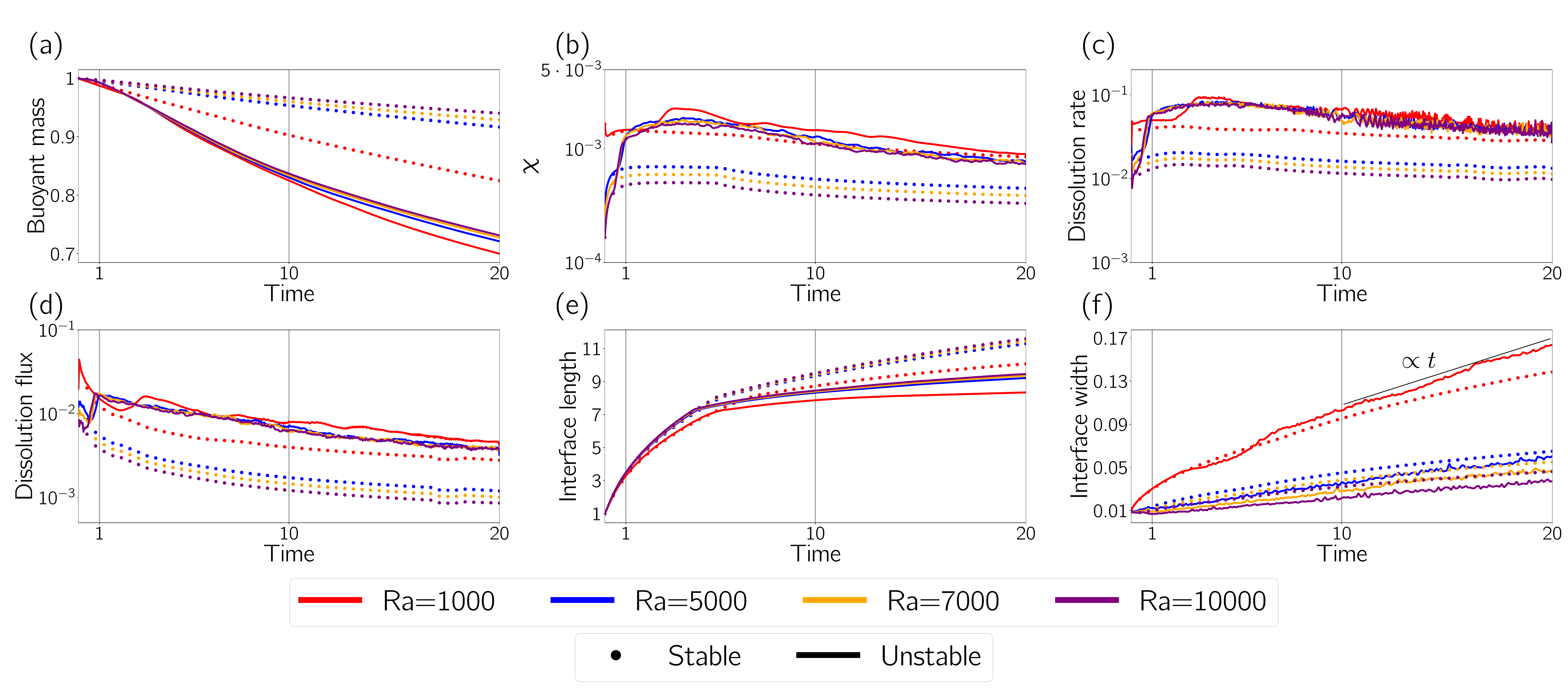}
        {\phantomsubcaption\label{fig:new_buoyantMass_homogeneous_stable_unstable}}
        {\phantomsubcaption\label{fig:new_scalarDissipationRate_homogeneous_stable_unstable}}
        {\phantomsubcaption\label{fig:new_dissolutionRate_homogeneous_stable_unstable}}
        {\phantomsubcaption\label{fig:new_dissolutionFlux_homogeneous_stable_unstable}}
        {\phantomsubcaption\label{fig:new_interfaceLength_homogeneous_stable_unstable}}
        {\phantomsubcaption\label{fig:new_interfaceWidth_homogeneous_stable_unstable}}
    \caption{Time evolution of the \subref{fig:new_buoyantMass_homogeneous_stable_unstable} buoyant mass, \subref{fig:new_scalarDissipationRate_homogeneous_stable_unstable} scalar dissipation rate, \subref{fig:new_dissolutionRate_homogeneous_stable_unstable} dissolution rate, \subref{fig:new_dissolutionFlux_homogeneous_stable_unstable} dissolution flux, \subref{fig:new_interfaceLength_homogeneous_stable_unstable} interface length, and \subref{fig:new_interfaceWidth_homogeneous_stable_unstable} interface width for the stable (dots) and unstable (solid line) case in a homogeneous medium. $\Rayleigh=1000, 5000, 7000, 10000$ is colored in red, blue, yellow and purple, respectively.}
    \label{fig:homo_stable_unstable}
\end{figure}

The time-evolution of the buoyant mass is shown in \autoref{fig:new_buoyantMass_homogeneous_stable_unstable}. It is monotonically decreasing because the domain can be considered semi-infinite with impervious boundaries and without sources as long as the gravity current is far from the right boundary. Moreover, we observe that the buoyant mass diminishes faster as $\Rayleigh$ decreases due to the increase of diffusion. The comparison of the stable and the unstable cases shows that the fingering instability favors dissolution.

In the stable case, the scalar dissipation rate decreases at initial times because the concentration gradient is reduced due to mixing (see \autoref{fig:new_scalarDissipationRate_homogeneous_stable_unstable} especially for $\Rayleigh = 1000$). Then, even though diffusion homogenizes the system, the scalar dissipation rate increases, precisely at the moment when the interface becomes a tilted line. At that moment, there is a competition between the decrease of the concentration gradient and the increase of the interface length. When the interface length grows faster than the concentration gradient decreases, the scalar dissipation rate increases (see \autoref{fig:new_scalarDissipationRate_homogeneous_stable_unstable} for $\Time > 0.3$). In contrast, when the concentration gradient decreases faster than the interface length increases, the scalar dissipation rate stops growing and decreases (see \autoref{fig:new_scalarDissipationRate_homogeneous_stable_unstable} at $\Time \approx 6$ and Sect. 4 of the supplementary material). 

In the unstable case, the scalar dissipation rate is initially constant because there is a balance between the stretching of the interface due to advection and the smoothing of the concentration gradient due to diffusion. Then, while the interface transitions from s-shaped into a tilted and straight line, the concentration gradient decreases in the interior of the interface and increases near the endpoints. Since the interface becomes longer, mixing occurs in a larger region, and the scalar dissipation rate increases smoothly (see \autoref{fig:new_scalarDissipationRate_homogeneous_stable_unstable}). When the first fingering instabilities appear, mixing is enhanced. Accordingly, the scalar dissipation rate increases abruptly (see \autoref{fig:new_scalarDissipationRate_homogeneous_stable_unstable} between $\Time \approx 0.5$ and $\Time \approx 2$) and the dissolution flux also increases but in a more moderate manner than $\ScalarDissipationRate$ since the interface is longer too (see \autoref{fig:new_dissolutionFlux_homogeneous_stable_unstable}).

When the fingers are present along all the interface, the mixing rate is consolidated. During this period, fingers appear and dissolve simultaneously and fluids are mixed at a constant rate. In this regime, the scalar dissipation rate is constant and attains its highest value (see \autoref{fig:new_scalarDissipationRate_homogeneous_stable_unstable} between $\Time \approx 2$ and $\Time \approx 7$). Finally,  when the plume's tail detaches from the bottom boundary, there is a change in the mixing rate (see \autoref{fig:new_scalarDissipationRate_homogeneous_stable_unstable}). Even though there are still fingers and fluids mix, the scalar dissipation rate decreases because the accumulation of dense fluid at the bottom restricts convective dissolution across a growing portion of the interface \citep{hidalgoDynamicsConvectiveDissolution2013}.

For the stable case, the dissolution rate and flux monotonically decrease with respect to $\Rayleigh$ (\autoref{fig:new_dissolutionRate_homogeneous_stable_unstable} and \autoref{fig:new_dissolutionFlux_homogeneous_stable_unstable}). The same can be observed for the unstable case accounting for the difference in the time for the onset of convection. In conclusion, although a high Rayleigh number accelerates the appearance of fingering instabilities, a low Rayleigh number favors mixing throughout all the time span.

The interface length is monotonically increasing for both the stable and unstable cases for the simulated time span (see \autoref{fig:new_interfaceLength_homogeneous_stable_unstable}). In both cases, we observe that the interface is longer for higher Rayleigh numbers because the convective term is more relevant. In the unstable case, the increasing trend of the interface length changes sharply when the plume detaches from the bottom boundary. Moreover, it detaches sooner from the bottom for higher Rayleigh numbers. 

Regarding the interface width, it increases monotonically with time, except for the unstable case at the onset of convection (see \autoref{fig:new_interfaceWidth_homogeneous_stable_unstable} and Fig. S7 in the supplementary material), when it is compressed, and its width is inversely proportional to $\Rayleigh$. For the unstable case, after the onset of convection, the interface width grows roughly linearly (see \autoref{fig:new_interfaceWidth_homogeneous_stable_unstable}). The difference between $\Rayleigh = 1000$ and $\Rayleigh \geq 5000$, is explained by the strong effect of diffusion and the late onset of convection for $\Rayleigh = 1000$.

\renewcommand{\figsize}{0.4}
\begin{figure}[t]
    \centering
    \includegraphics[width=\textwidth]{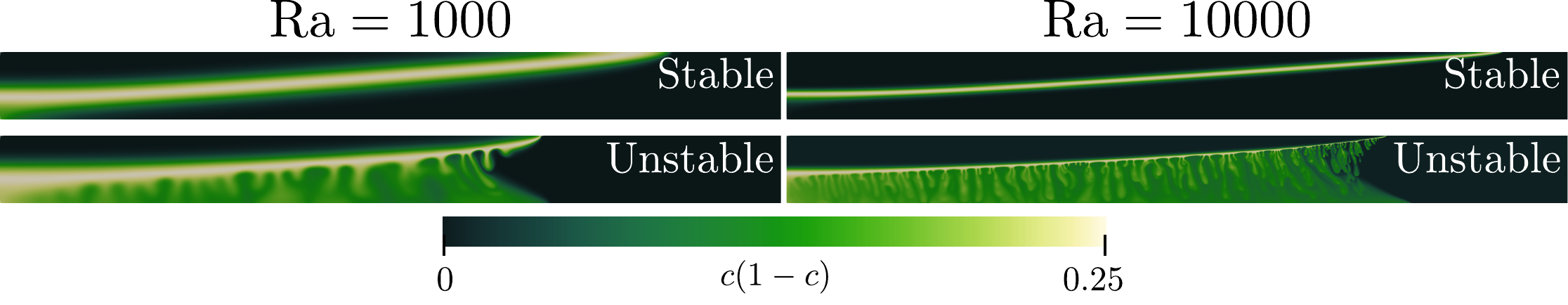}
    \caption{Snapshots of the field $\Concentration \left( 1 - \Concentration \right)$ for the stable (top) and unstable (bottom) case in a homogeneous medium for $\Rayleigh = 1000$ (left) and $\Rayleigh = 10000$ (right) at time $\Time = 15$. The domain extends to $\Spacex = 20$, but only $0 \leq \Spacex \leq 12$ is shown here.}
    \label{fig:new_comparisonGravityCurrentHomo_interfaceWidth}
\end{figure}

For $\Rayleigh = 1000$, we observe that the interface is wider for the unstable case than for the stable case, but this trend is reversed for $\Rayleigh \geq 5000$. Consider the simulation snapshots in \autoref{fig:new_comparisonGravityCurrentHomo_interfaceWidth}. The stable case presents a uniform interface width. In the unstable case, however, the upward-flowing brine compresses the interface, which is also stretched due to a high lateral velocity. At the plume's tail the interface becomes very wide after it detaches from the bottom and substantially contributes to the average value (see \autoref{fig:new_comparisonGravityCurrentHomo_interfaceWidth}). For $\Rayleigh \geq 5000$, this wide region compensates for the thin part of the rest of the interface and the average value becomes similar to the stable case (see \autoref{fig:new_interfaceWidth_homogeneous_stable_unstable}). In contrast, for $\Rayleigh = 1000$, the average value is dominated by the width at the plume's tail, which is wider for the unstable case than for the stable case (\autoref{fig:new_comparisonGravityCurrentHomo_interfaceWidth}), and it is not compensated by the width away from the plume's tail because the interface does not undergo a high compression and stretching. Thus, the overall interface width for the unstable case is larger, as seen in \autoref{fig:new_interfaceWidth_homogeneous_stable_unstable} for $\Time \geq 5$.
%
%
\subsection{Gravity currents in heterogeneous isotropic porous media}
\label{sec:new_isotropic}

\renewcommand{\figsize}{0.4}
\begin{figure}[t]
    \centering
    \includegraphics[width=\textwidth]{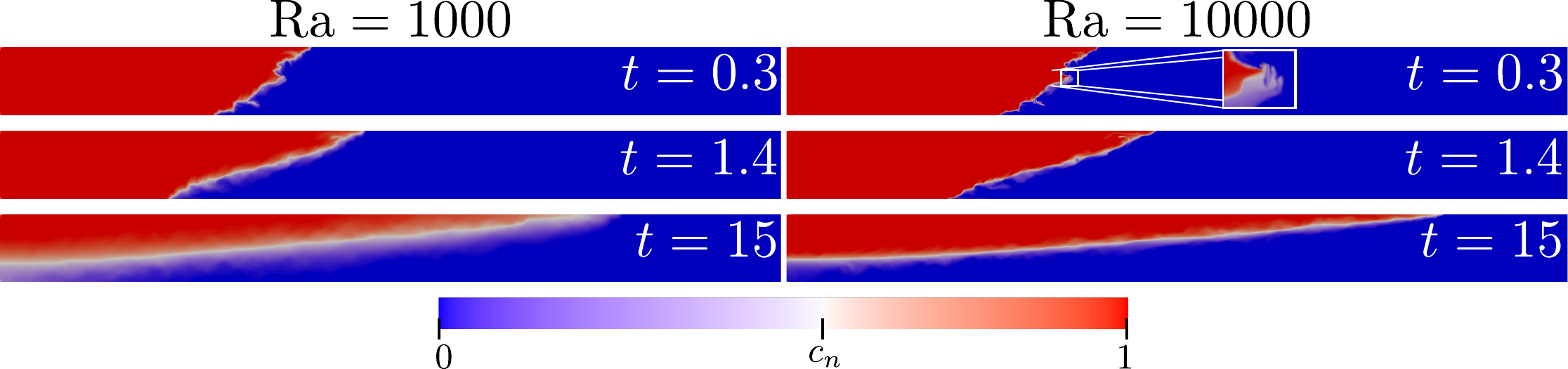}
    \caption{Snapshots of the concentration field for the stable case in a realization of a heterogeneous isotropic medium with correlation length $\CorrLenx = \CorrLenz = 0.05$ and variance $\Variance = 1$ for $\Rayleigh = 1000$ (left) and $\Rayleigh = 10000$ (right) at times $\Time = 0.3$ (top), $\Time = 1.4$ (middle), and $\Time = 15$ (bottom). The domain extends to $\Spacex = 20$, but only $0 \leq \Spacex \leq 12$ is shown here.}
    \label{fig:comparisonGravityCurrentHeteroStable}
\end{figure}

\renewcommand{\figsize}{0.4}
\begin{figure}[t]
    \centering
    \includegraphics[width=\textwidth]{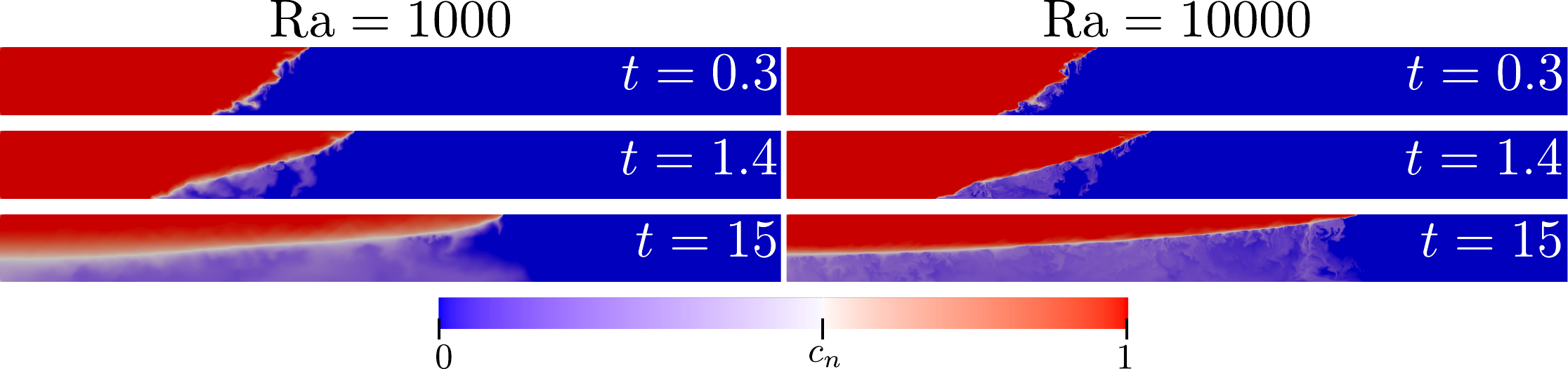}
    \caption{Snapshots of the concentration field for the unstable case in a realization of a heterogeneous isotropic medium with correlation length $\CorrLenx = \CorrLenz = 0.05$ and variance $\Variance = 1$ for $\Rayleigh = 1000$ (left) and $\Rayleigh = 10000$ (right) at times $\Time = 0.3$ (top), $\Time = 1.4$ (middle), and $\Time = 15$ (bottom). The domain extends to $\Spacex = 20$, but only $0 \leq \Spacex \leq 12$ is shown here.}
    \label{fig:comparisonGravityCurrentHeteroUnstable}
\end{figure}

In this section, we study the effect of the heterogeneity in the plume dynamics and mixing for the stable and unstable cases. We consider permeability fields with $\CorrLenx = \CorrLenz = 0.05$ and variance $\Variance = 1$.

In a heterogeneous medium, heterogeneity stirs the fluid affecting mixing. At initial times, the interface becomes rapidly tortuous due to the heterogeneity and finger formation is triggered due to two mechanisms. On the one hand, only in the unstable case (\autoref{fig:comparisonGravityCurrentHeteroUnstable}), fingering appears because of the unstable density stratification created by the non-monotonic density law. On the other hand, the contrast between low and high permeability regions favors upward finger formation. Blobs form when the light fluid gets trapped in low-permeability regions surrounded by dense fluid and flows upward creating fingering structures (see \autoref{fig:comparisonGravityCurrentHeteroStable} and Fig. S9 in the supplementary material). This phenomenon affects both the stable and unstable cases (see Movie M2 in the supplementary material) and is exaggerated by medium anisotropy (see Movie M3 in the supplementary material).

As fluids mix, blobs disappear and the interface can be approximated by a tilted straight line, yet it accommodates the medium heterogeneity. In the stable case, the interface becomes wider due to diffusion and, on average, there are less and smaller large concentration gradient regions than in the unstable case (\autoref{fig:comparisonGravityCurrentHeteroStable}). Thus, mixing is diminished. In the unstable case, the dense fluid sinks to the bottom following high-permeability paths (see \autoref{fig:comparisonGravityCurrentHeteroUnstable} for $\Time = 1.4$ and Fig. S10 in the supplementary material). Nevertheless, fingers that have formed in a high permeability region slow down when they reach a low-permeable zone. That is, low permeability regions prevent penetration of fingers and favor dissipation due to diffusion. We have previously observed this phenomenon in the homogeneous case (see \autoref{sec:new_homogeneous}). In that case, the layer of dense fluid accumulates at the bottom of the domain and hinders mixing. In contrast, in the heterogeneous case, the dense fluid accumulates in each low permeability region and hinders mixing locally. Finally, as the plume migrates and its tail detaches from the bottom, it accumulates below the plume. Thus, fingers tend to proliferate close to the nose of the plume, where the concentration gradient is larger (see \autoref{fig:comparisonGravityCurrentHeteroUnstable} for $\Time = 15$).

\renewcommand{\figsize}{0.32}
\begin{figure}[t]
    \centering
    \includegraphics[width=\linewidth]{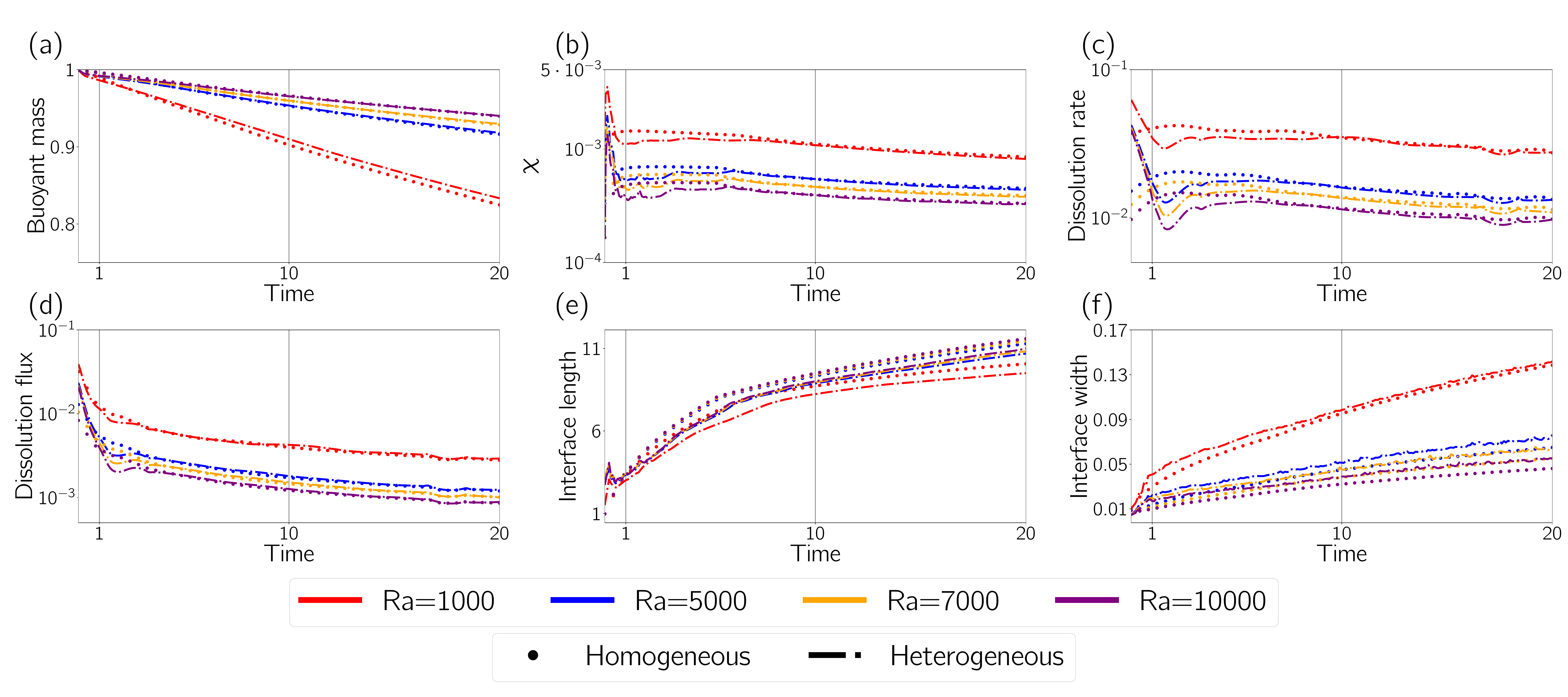}
    {\phantomsubcaption\label{fig:new_buoyantMass_homogeneous_heterogeneous_stable}}
    {\phantomsubcaption\label{fig:new_scalarDissipationRate_homogeneous_heterogeneous_stable}}
    {\phantomsubcaption\label{fig:new_dissolutionRate_homogeneous_heterogeneous_stable}}
    {\phantomsubcaption\label{fig:new_dissolutionFlux_homogeneous_heterogeneous_stable}}
    {\phantomsubcaption\label{fig:new_interfaceLength_homogeneous_heterogeneous_stable}}
    {\phantomsubcaption\label{fig:new_interfaceWidth_homogeneous_heterogeneous_aniso_stable}}
    \caption{Time evolution of the \subref{fig:new_buoyantMass_homogeneous_heterogeneous_stable} buoyant mass, \subref{fig:new_scalarDissipationRate_homogeneous_heterogeneous_stable} scalar dissipation rate, \subref{fig:new_dissolutionRate_homogeneous_heterogeneous_stable} dissolution rate, \subref{fig:new_dissolutionFlux_homogeneous_heterogeneous_stable} dissolution flux, \subref{fig:new_interfaceLength_homogeneous_heterogeneous_stable} interface length, and \subref{fig:new_interfaceWidth_homogeneous_heterogeneous_aniso_stable} interface width for the stable case in a homogeneous (dots) and heterogeneous (dash-dot line) isotropic medium with correlation length $\CorrLenx = \CorrLenz = 0.05$ and variance $\Variance = 1$. $\Rayleigh=1000, 5000, 7000, 10000$ is colored in red, blue, yellow and purple, respectively.}
    \label{fig:homo_hetero_stable}
\end{figure}

The stable case shows the effect of the heterogeneity in the plume migration and mixing in the absence of convective dissolution (\autoref{fig:homo_hetero_stable} and Fig. S3 in the supplementary material). In heterogeneous media, the concentration gradient is high not only at the nose and tail of the plume, but also at regions where there exists a large permeability contrast. Since the light fluid may be trapped in low-permeability regions, these disconnected components also contribute to the interface length. The combination of more regions where the concentration gradient is large and a longer interface makes the scalar dissipation rate increase in the heterogeneous case (see \autoref{fig:new_scalarDissipationRate_homogeneous_heterogeneous_stable} for $\Time < 0.1$).  Thus, it is during this time period that we observe that more mass is dissolved for the heterogeneous than for the homogeneous case (\autoref{fig:new_buoyantMass_homogeneous_heterogeneous_stable}).
As time goes on, the interface becomes wider due to diffusion and, on average, there are less and smaller regions where the concentration gradient is large. The trapped fluid dissolves and the interface length decreases (see \autoref{fig:new_interfaceLength_homogeneous_heterogeneous_stable} between $\Time = 10^{-1}$ and $\Time = 1$). Consequently, the scalar dissipation rate decreases (see \autoref{fig:new_scalarDissipationRate_homogeneous_heterogeneous_stable} until $\Time \approx 1$). From this point on, the interface is almost a tilted straight line, and, although low (high) permeability regions may prevent (ease) migration, the homogeneous and heterogeneous problems behave in a similar manner.

\renewcommand{\figsize}{0.32}
\begin{figure}[t]
    \centering
    \includegraphics[width=\linewidth]{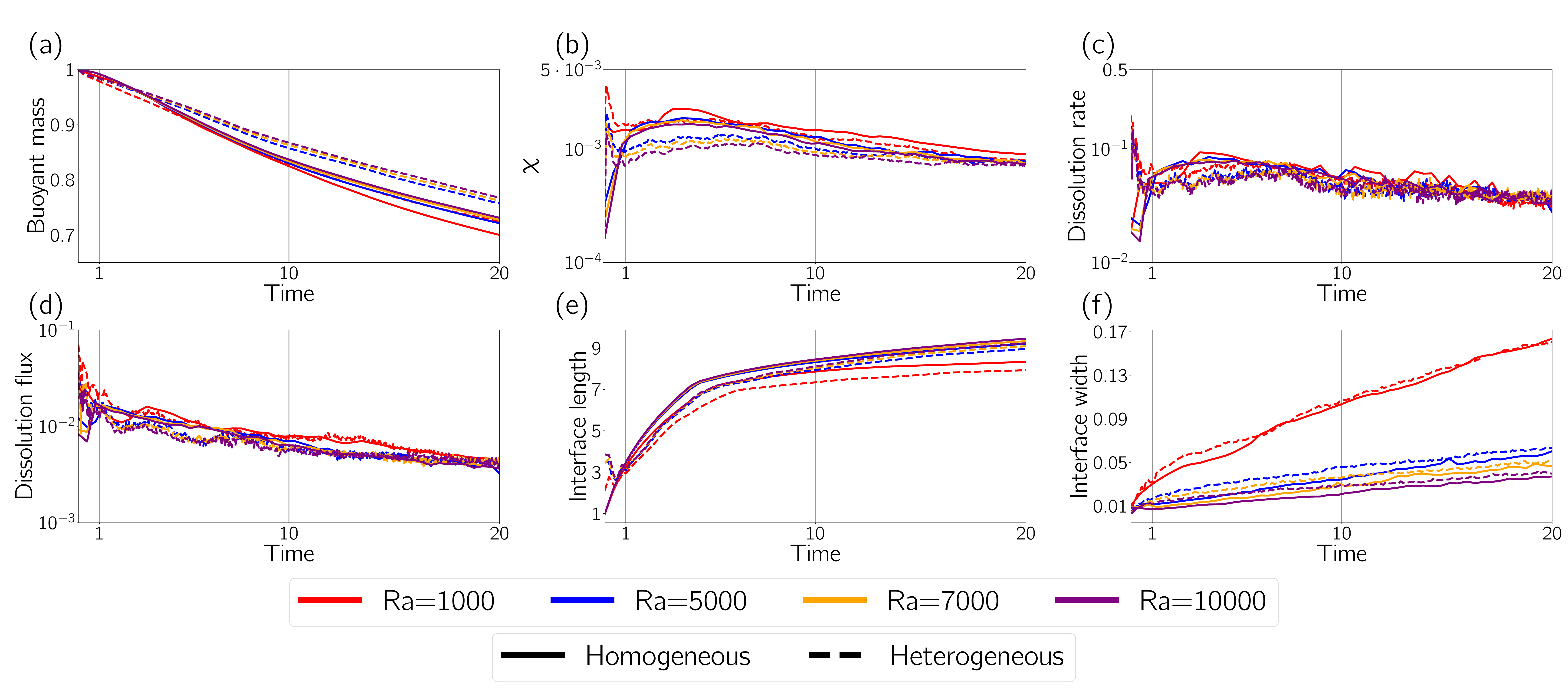}
    {\phantomsubcaption\label{fig:new_buoyantMass_homogeneous_heterogeneous_iso_unstable}}
    {\phantomsubcaption\label{fig:new_scalarDissipationRate_homogeneous_heterogeneous_iso_unstable}}
    {\phantomsubcaption\label{fig:new_dissolutionRate_homogeneous_heterogeneous_iso_unstable}}
    {\phantomsubcaption\label{fig:new_dissolutionFlux_homogeneous_heterogeneous_iso_unstable}}
    {\phantomsubcaption\label{fig:new_interfaceLength_homogeneous_heterogeneous_iso_unstable}}
    {\phantomsubcaption\label{fig:new_interfaceWidth_homogeneous_heterogeneous_iso_unstable}}
    \caption{Time evolution of the \subref{fig:new_buoyantMass_homogeneous_heterogeneous_iso_unstable} buoyant mass, \subref{fig:new_scalarDissipationRate_homogeneous_heterogeneous_iso_unstable} scalar dissipation rate, \subref{fig:new_dissolutionRate_homogeneous_heterogeneous_iso_unstable} dissolution rate, \subref{fig:new_dissolutionFlux_homogeneous_heterogeneous_iso_unstable} dissolution flux, \subref{fig:new_interfaceLength_homogeneous_heterogeneous_iso_unstable} interface length, and \subref{fig:new_interfaceWidth_homogeneous_heterogeneous_iso_unstable} interface width for the unstable case in a homogeneous (solid line) and heterogeneous (dashed line) isotropic medium with correlation length $\CorrLenx = \CorrLenz = 0.05$ and variance $\Variance = 1$. $\Rayleigh=1000, 5000, 7000, 10000$ is colored in red, blue, yellow and purple, respectively.}
    \label{fig:homo_hetero_unstable}
\end{figure}

The unstable case shows the interaction of the heterogeneity with the fingering instabilities (\autoref{fig:homo_hetero_unstable} and Sect. 5 of the supplementary material). The onset of convection is accelerated by heterogeneity and more mass is dissolved at early times (\autoref{fig:new_buoyantMass_homogeneous_heterogeneous_iso_unstable}). This is reflected in the increase of the scalar dissipation rate (\autoref{fig:new_scalarDissipationRate_homogeneous_heterogeneous_iso_unstable}) and the sudden increase of the interface length (\autoref{fig:new_interfaceLength_homogeneous_heterogeneous_iso_unstable}).
Then, as the fluid is trapped in low-permeability regions ($10^{-1} \lessapprox \Time  \lessapprox 1$), it gets mixed, the interface becomes shorter, and $\ScalarDissipationRate$ and the dissolution flux decrease.

The scalar dissipation rate displays an oscillating behavior with a maximum value at $\Time \approx 0.1$. After that, it declines and increases again at $\Time \approx 1$. This behavior is caused by the intermittent attenuation of the fingering instability by low permeability layers and the creation of new fingers at the nose of the gravity current. In the homogeneous case, fingers are well-defined and last longer. Consequently, more buoyant mass is dissolved in the homogeneous case not only after the onset of convection but also globally (\autoref{fig:new_buoyantMass_homogeneous_heterogeneous_iso_unstable}). There is also a faster growth of the interface and a higher dissolution rate and flux with respect to the heterogeneous case. The interface, however, is more compressed because it does not experience the dispersive effect of heterogeneity. 

Finally, in both homogeneous and heterogeneous media, as convection is shutdown by the dense mixture that accumulates below the gravity current, the scalar dissipation rate, dissolution rate and flux decrease in a similar manner. 

The effect of $\Rayleigh$ is similar for both stable and unstable cases, regardless of heterogeneity, with a wider interface and a higher $\ScalarDissipationRate$ as $\Rayleigh$ decreases. In heterogeneous media, the separation among different $\Rayleigh$ remains clearly distinguishable during the whole simulated time for $\ScalarDissipationRate$, whereas in the homogeneous medium values are very similar after the onset of convection. This suggests that the effective $\Rayleigh$ decreases with heterogeneity and convective mixing remains far from the Rayleigh-independent asymptotic value.

\renewcommand{\figsize}{0.4}
\begin{figure}[t]
    \centering
    \includegraphics[width=\textwidth]{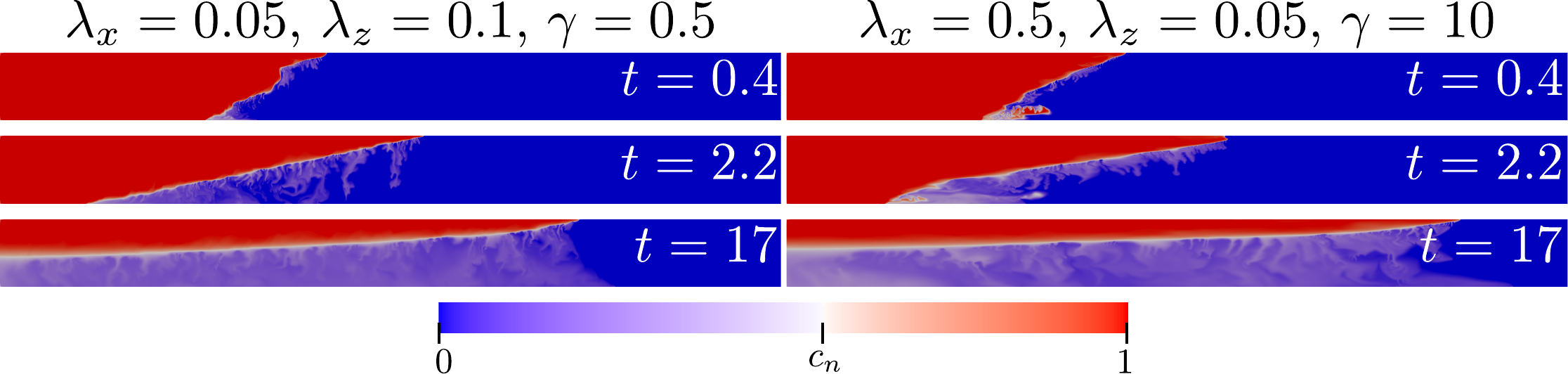}
    \caption{Snapshots of the concentration field for the unstable case for $\Rayleigh = 5000$ in a heterogeneous porous medium with variance $\Variance = 1$ and correlation-length anisotropy ratios $\RatioCorrLen = 0.5$ (left) and $\RatioCorrLen = 10$ (right) at times $\Time = 0.4$ (top), $\Time = 2.2$ (middle), and $\Time = 17$ (bottom). The domain extends to $\Spacex = 20$, but only $0 \leq \Spacex \leq 12$ is shown here.}
    \label{fig:comparisonGravityCurrentHeteroAnisotropy}
\end{figure}

%
%
\subsection{Impact of permeability correlation length}
\label{sec:anisotropy}

In this section, we study the effect of different correlation-length anisotropy values (see \autoref{sec:generationPermeability}). Specifically, we consider four scenarios with correlation-length anisotropy ratios $\RatioCorrLen := \frac{\CorrLenx}{\CorrLenz} =$ 0.5, 1, 2, and 10. We fix here $\Rayleigh = 5000$ and $\Variance = 1$. (see Fig. \ref{fig:comparisonGravityCurrentHeteroAnisotropy} and Movie M4 in supplementary material).

\renewcommand{\figsize}{0.32}
\begin{figure}[t]
    \centering
    \includegraphics[width=\linewidth]{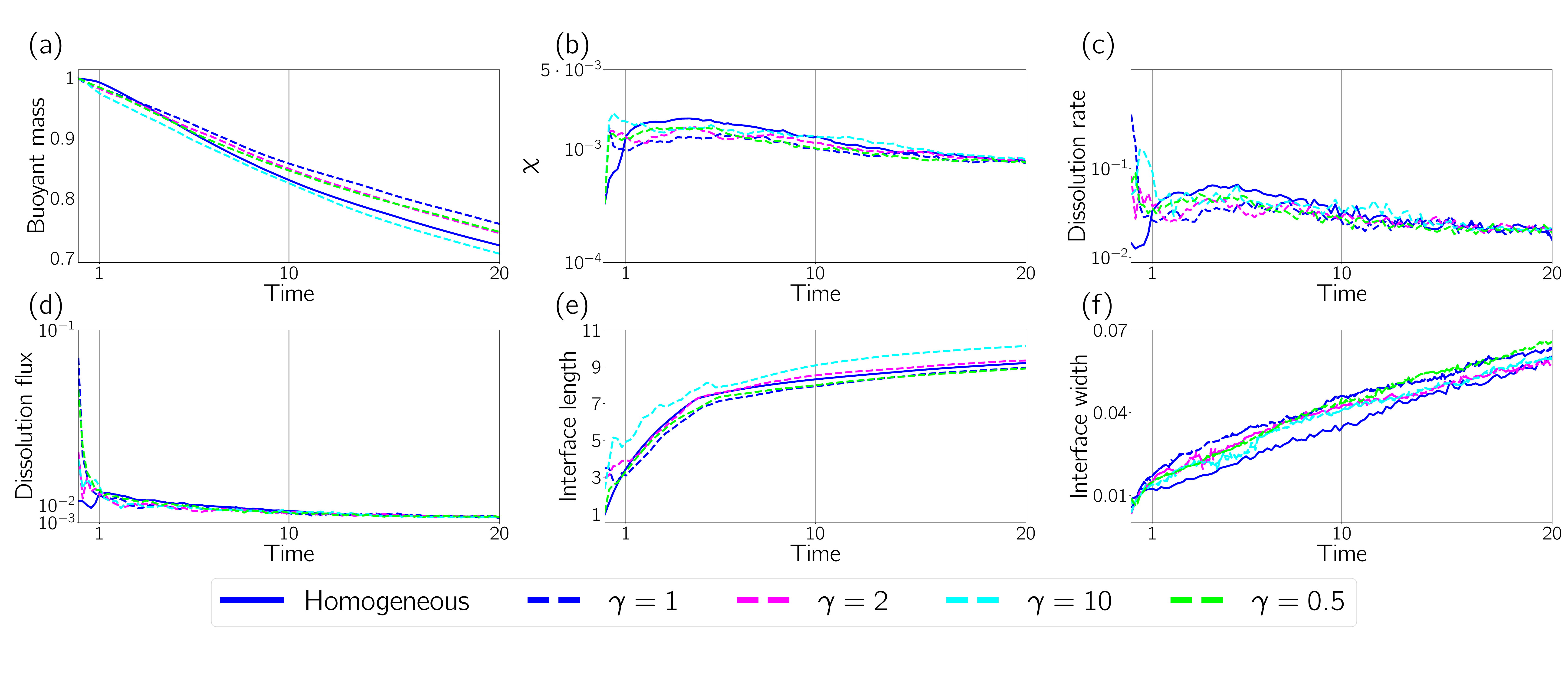}
    {\phantomsubcaption\label{fig:buoMassHeteroAnisotropy}}
    {\phantomsubcaption\label{fig:scalarDissipationRateHeteroAnisotropy}}
    {\phantomsubcaption\label{fig:dissRateHeteroAnisotropy}}
    {\phantomsubcaption\label{fig:dissFluxHeteroAnisotropy}}
    {\phantomsubcaption\label{fig:interfaceHeteroAnisotropy}}
    {\phantomsubcaption\label{fig:widthHeteroAnisotropy}}
    \caption{Time evolution of the \subref{fig:buoMassHeteroAnisotropy} buoyant mass, \subref{fig:scalarDissipationRateHeteroAnisotropy} scalar dissipation rate, \subref{fig:dissRateHeteroAnisotropy} dissolution rate, \subref{fig:dissFluxHeteroAnisotropy} dissolution flux, \subref{fig:interfaceHeteroAnisotropy} interface length, and \subref{fig:widthHeteroAnisotropy} interface width for different values of the correlation-length anisotropy ratio. $\RatioCorrLen = 0.5$, 1, 2, 10 is colored in lime, blue, magenta and cyan, respectively. The solid curve corresponds to the homogeneous case, and the dashed lines to heterogeneous permeability cases.}
    \label{fig:HeteroAnisotropy}
\end{figure}

For the vertically stratified case ($\RatioCorrLen < 1$), fingering instabilities appear throughout the interface and sink to the bottom of the domain (see left panel of \autoref{fig:comparisonGravityCurrentHeteroAnisotropy}). For the horizontally stratified cases ($\RatioCorrLen > 1$), we observe that the lighter fluid gets trapped in the low-permeable layers (see top-right panel of \autoref{fig:comparisonGravityCurrentHeteroAnisotropy}), and fingers become rapidly suppressed by the fluid mixture trapped between the layers (see middle-right panel of \autoref{fig:comparisonGravityCurrentHeteroAnisotropy}). Moreover, the plume migrates farther for $\RatioCorrLen > 1$ than for $\RatioCorrLen < 1$ (compare bottom panels of \autoref{fig:comparisonGravityCurrentHeteroAnisotropy} and nose position in Fig. S5 in supplementary material).

For the horizontally stratified cases ($\RatioCorrLen > 1$), the buoyant mass decreases as $\RatioCorrLen$ increases (\autoref{fig:buoMassHeteroAnisotropy}).
All heterogeneous cases dissolve less mass than the homogeneous case except for the $\RatioCorrLen = 10$ case, in which the fluid trapped in the horizontal low permeability layers increases the dissolution (see the trapped regions below the gravity current in the panels on the right of \autoref{fig:comparisonGravityCurrentHeteroAnisotropy}). Although the heterogeneous cases dissolve more mass than the homogeneous case at early times because of the early onset of convection (see \autoref{fig:buoMassHeteroAnisotropy} and \autoref{fig:scalarDissipationRateHeteroAnisotropy}), the barrier effect of the horizontal stratification of the permeability, that suppresses finger formation, reduces rapidly the dissolution rate and the dissolution flux (\autoref{fig:dissRateHeteroAnisotropy} and \autoref{fig:dissFluxHeteroAnisotropy}). This results in an overall lower dissolved mass than in the homogeneous case.

The interface length (\autoref{fig:interfaceHeteroAnisotropy}) and the maximum $\ScalarDissipationRate$ (\autoref{fig:scalarDissipationRateHeteroAnisotropy}) display the same trend. The $\RatioCorrLen = 10$ case, again, presents the longest interface because of the contribution of the trapped fluid blobs, which also give the interface length an oscillatory behavior. Anisotropic cases have a wider interface than the homogeneous case, which is consistent with the barrier effect and a reduction of the concentration gradient driving dissolution.

The vertically stratified case ($\RatioCorrLen = 0.5$) has a late onset of convection but still dissolves more mass than the homogeneous case at initial times. The vertical stratification acts as a vertical barrier for the migration of the gravity current as can be seen in the slower advancement of the plume's nose with respect to the $\RatioCorrLen = 10$ case (see \autoref{fig:comparisonGravityCurrentHeteroAnisotropy} and Fig. S5 in supplementary material). This hindered advancement of the gravity current causes the scalar dissipation rate, dissolution flux and dissolution rate to have an intermittent behavior. Vertical stratification also favors the formation of fingers \citep{benhammadiEffectPermeabilityHeterogeneity2025a}, which widens the interface and reduces the dissolution flux. The fast migration of fingers to the bottom of the domain also accelerates the starting of the shutdown regime signaled by the steady decrease of $\ScalarDissipationRate$.

These observations reflect that the interaction between the size of the fingering instability and the size of the correlation-length anisotropy of the permeability field (see Fig. S10 in the supplementary material) affects the behavior of the gravity current. A similar behavior was observed by \citep{benhammadiExperimentalNumericalStudy2025} for a horizontal interface.
%
%
\subsection{Impact of the variance of the permeability field}
\label{sec:variance}

\renewcommand{\figsize}{0.4}
\begin{figure}[t]
    \centering
    \includegraphics[width=\textwidth]{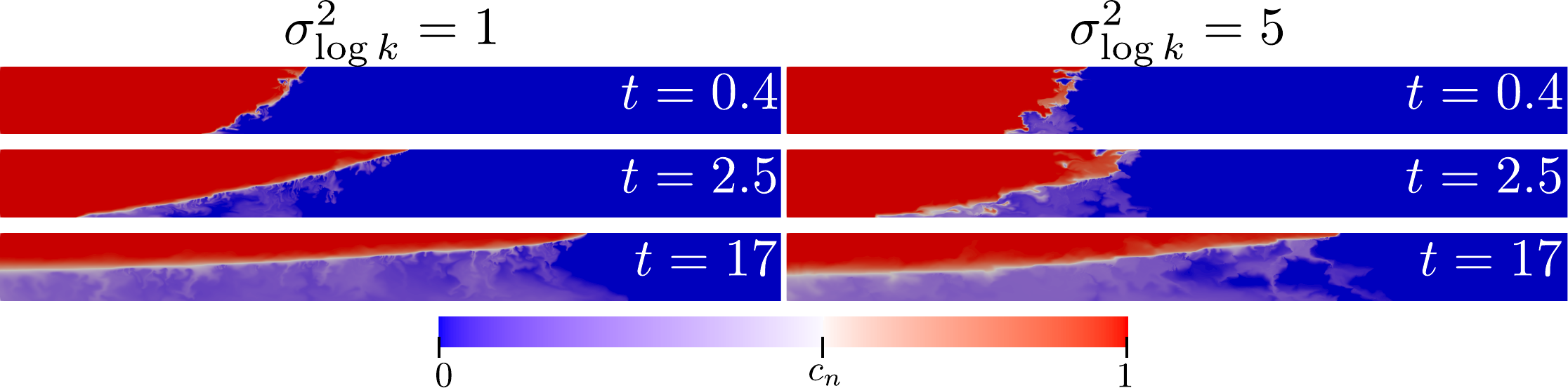}
    \caption{Snapshots of the concentration field for the unstable case for $\Rayleigh = 5000$ in a heterogeneous porous medium with correlation length $\CorrLenx = 0.1$, $\CorrLenz = 0.05$, and variance $\Variance = 1$ (left) and $\Variance = 5$ (right) at times $\Time = 0.4$ (top), $\Time = 2.5$ (middle), and $\Time = 17$ (bottom). The domain extends to $\Spacex = 20$, but only $0 \leq \Spacex \leq 12$ is shown here.}
    \label{fig:comparisonGravityCurrentHeteroVariance}
\end{figure}

In this section, we study the effect of the variance of the log-normal distribution of the permeability field. Specifically, we compare $\Variance = 1$, 3, and 5, for $\Rayleigh = 1000$, 5000, 7000, and 10000. Since on-site porous media tend to have a higher correlation in the horizontal direction \citep{Turcke1996}, we consider $\RatioCorrLen = 2$ with $\CorrLenx = 0.1$. Snapshots of the simulation for three different times of the cases with $\Variance=1, 5$ are shown in \autoref{fig:comparisonGravityCurrentHeteroVariance} (see Movie M5 in supplementary material for the whole animation).

\renewcommand{\figsize}{0.32}
\begin{figure}[t]
    \centering
    \includegraphics[width=\linewidth]{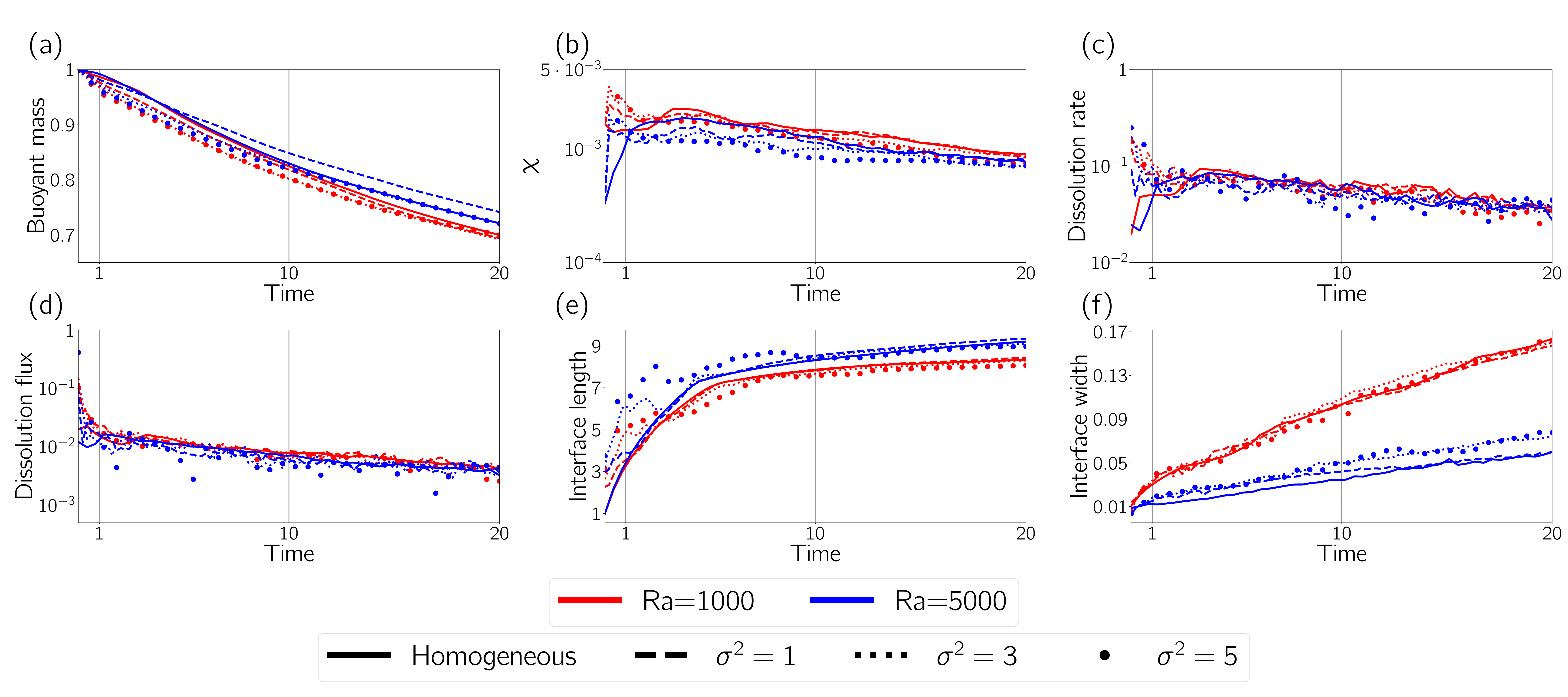}
    {\phantomsubcaption\label{fig:new_buoyantMass_homogeneous_heterogeneous_variance}}
    {\phantomsubcaption\label{fig:new_scalarDissipationRate_homogeneous_heterogeneous_variance}}
    {\phantomsubcaption\label{fig:new_dissolutionRate_homogeneous_heterogeneous_variance}}
    {\phantomsubcaption\label{fig:new_dissolutionFlux_homogeneous_heterogeneous_variance}}
    {\phantomsubcaption\label{fig:new_interfaceLength_homogeneous_heterogeneous_variance}}
    {\phantomsubcaption\label{fig:new_interfaceWidth_homogeneous_heterogeneous_variance}}
    \caption{Time evolution of the \subref{fig:new_buoyantMass_homogeneous_heterogeneous_variance} buoyant mass, \subref{fig:new_scalarDissipationRate_homogeneous_heterogeneous_variance} scalar dissipation rate, \subref{fig:new_dissolutionRate_homogeneous_heterogeneous_variance} dissolution rate, \subref{fig:new_dissolutionFlux_homogeneous_heterogeneous_variance} dissolution flux, \subref{fig:new_interfaceLength_homogeneous_heterogeneous_variance} interface length, and \subref{fig:new_interfaceWidth_homogeneous_heterogeneous_variance} interface width for different values of the variance of the log-normal distribution characterizing the permeability field. Variance $\Variance = 1$, 3, 5 are represented with dashed, dotted and circle lines, respectively. The solid curve corresponds to the unstable homogeneous case. Rayleigh number $\Rayleigh=1000, 5000$ is colored in red and blue, respectively.}
    \label{fig:homo_hetero_unstable_variance}
\end{figure}

The time-evolution of the buoyant mass shows, consistent with the anisotropic cases with $\RatioCorrLen > 1$, that, for high $\Rayleigh$, dissolution is always less efficient than in homogeneous media (\autoref{fig:new_buoyantMass_homogeneous_heterogeneous_variance} and Fig. S11 in the supplementary material for the $\Rayleigh=7000$ and 10000 cases). In heterogeneous media, the efficiency increases with $\Variance$. This is caused by the acceleration of the onset of convection and the initial increase in the interface length together with a high gradient across the interface (\autoref{fig:new_scalarDissipationRate_homogeneous_heterogeneous_variance}-\subref{fig:new_interfaceLength_homogeneous_heterogeneous_variance}). However, as the gravity current evolves, the interface widens reducing the concentration gradient driving dissolution and the dissolution flux and dissolution rate decrease (\autoref{fig:new_dissolutionRate_homogeneous_heterogeneous_variance}-\subref{fig:new_interfaceWidth_homogeneous_heterogeneous_variance}) below the level of the homogeneous case. The relative increase of the interface width is more pronounced for high $\Variance$, revealing the dispersive effect of heterogeneity. The resulting decrease in the dissolution efficiency is reflected in the slow but steady decay of $\ScalarDissipationRate$ shortly after the onset of convection (\autoref{fig:new_scalarDissipationRate_homogeneous_heterogeneous_variance}).

For low $\Rayleigh$, even though convection is not as strong as in the high $\Rayleigh$ cases, diffusion across the interface is strong enough to dissolve a great quantity of fluid mass. The effect of heterogeneity on the interface width is relatively small because the size of the fingers is much larger than the correlation lengths $\CorrLenx$ and $\CorrLenz$, and, therefore, the interaction between the permeability field and the instability is weak \citep{benhammadiEffectPermeabilityHeterogeneity2025a}.

%
%
\section{Summary and conclusions}
\label{sec:conclusions}

We investigated the impact of the heterogeneity of the permeability field on the mixing, migration, and spreading of gravity currents in porous media by means of high-fidelity numerical simulations. We considered a stable case with a linear density law and an unstable case with a non-monotonic density law mimicking a \COO{}-brine system. Heterogeneity was modeled using log-normal permeability fields of different variances and correlation lengths. Although we only considered the average of two realizations of the permeability field and statistically converged trends cannot be extracted, our results show that observables such as the buoyant mass, scalar dissipation rate, dissolution flux and rate, and interface length and width are strongly influenced by heterogeneity.

The comparison between the stable and unstable cases in both homogeneous and heterogeneous scenarios confirms the enhancement of the dissolution caused by the fingering instability. After the onset of convection, the unstable case presents a shorter but narrower interface between the fluids than the stable case and sustains a higher mixing and dissolution over the life of the gravity current. However, dissolution slows the gravity current and the plume's nose advances further in the stable case.

Heterogeneity affects each case in a different way. In the stable case, only at early times $(t<1)$, the dispersive effect of the heterogeneity widens the interface and favors mixing and dissolution. Therefore, it decreases the speed of the gravity current and the plume's nose lags behind the homogeneous scenario. At late times, however, both scenarios behave similarly. In the unstable case, the early time behavior is similar to the stable case and heterogeneity increases dissolution. Heterogeneity accelerates the onset of convection and this causes a rapid increase in dissolution and interface length. At late times, during the convection-dominated and shutdown regimes, the fingering instability is not able to counteract the dispersive effect of heterogeneity. The heterogeneous scenario has a wider and shorter interface than the homogeneous scenario and dissolution and mixing are not as efficient. In the stable and unstable cases, the trapping of buoyant fluid in low permeability regions during the initial rotation of the interface leads to the formation of fluid blobs that rapidly dissolve, creating an upward fingering instability in the stable case. These blobs increase the dissolution intermittently and deform the interface between the fluid favoring mixing.

The correlation structure of the permeability field plays an important role. Horizontally stratified media ($\RatioCorrLen > 1$) create a barrier effect for the sinking fingers that shut down convection locally and reduce the dissolution flux at the same time that widen the interface. Vertically stratified media ($\RatioCorrLen < 1$) present a barrier effect to the advancement of the current, but favor downward migration of the fingers. As a result, the scalar dissipation rate and dissolution display an intermittent behavior as the gravity current crosses the permeability barriers and the convection shutdown happens earlier than in the other isotropic and anisotropic cases.

The efficiency of dissolution is enhanced by the strength of the heterogeneity, i.e., $\Variance$. However, high values of $\Variance$ are needed to obtain similar dissolution efficiency to the homogeneous case when $\Rayleigh$ is high. The early onset of convection and the high dissolution flux observed in the heterogeneous cases have a short time span that makes the heterogeneous cases underperform the homogeneous case, except for small $\Rayleigh$.

In conclusion, the interaction between the size of the instabilities and the correlation structure of the permeability field determines the behavior of the gravity current. Although in all cases heterogeneity affects the shape and behavior of the gravity current, the dispersive and barrier effects of the permeability field are more important when the sizes are similar (high $\Rayleigh$ in this study) and, therefore, the dissolution of the fluid and the advance of the gravity current are hindered compared to the homogeneous case. Only when the interaction between the instability is weak (low $\Rayleigh$) does the early onset of convection and interface elongation caused by heterogeneity contribute to higher dissolution than in the homogeneous case. This suggests that the combination of the strength and size of the instability ($\Rayleigh$) and heterogeneity ($\RatioCorrLen, \Variance$) is a determinant factor of dissolution efficiency, which deserves further investigation.

These findings highlight the fundamental role of permeability heterogeneity in the migration and dissolution of gravity currents. They indicate the necessity to include these effects in sharp interface and vertically integrated models to accurately reproduce and estimate the evolution of \COO{} injected plumes or seawater intrusion wedges and capture the mixing and dissolution processes.

%
%
\appendix
\section{Dimensionless form of the governing equations}
\label{sec:dimensionlessProcedure}

In this section, we detail the nondimensionalization procedure of the governing equations \citep{riazOnsetConvectionGravitationally2006}. We describe dimensional quantities with a letter without tilde, and add a tilde for dimensionless quantities.

We model a semi-infinite aquifer with a two-dimensional rectangular domain $\Domain$ in the $x$-$z$ plane, with dimensions $\Domain = \left[ 0, \Lx \right] \times \left[ 0, \Lz \right]$. The aquifer is tilted by an angle $\Angle$ relative to the horizontal. The coordinate axes are aligned with the domain, i.e., a point $\bmSpace = \left( \Spacex, \Spacez \right) \in \Domain$ is given by $\bmSpace = \Spacex \UnitVecx + \Spacez \UnitVecz$. Thus, in this frame, the gravity vector $\GravityVec$ is expressed as $\GravityVec = \GravityCt \GravityUnit$, where $\GravityUnit = \left( - \sin \Angle, - \cos \Angle \right)$.

Fluid mass conservation reads
\[
\frac{\partial \Density}{\partial \Time} + \Div \left( \Density \DarcyVeloc \right) = \Source,
\]
where $\Density \equiv \Density \SpaceTime$ and $\DarcyVeloc \equiv \DarcyVeloc \SpaceTime$ denote the fluid density and the Darcy velocity, respectively. The term $\Source \equiv \Source \SpaceTime$ accounts for sources or sinks. Herein, we consider $\Source = 0$.

The Darcy-scale flow equation is given by the Darcy equation
\begin{equation}
	\DarcyVeloc = - \frac{\PermTensor}{\Viscosity} \left( \Grad \Pressure - \Density \GravityVec \right),
	\label{app_eq:DarcyDim}
\end{equation}
where $\PermTensor \equiv \PermTensor \left( \bmSpace \right)$ is the permeability tensor, $\Viscosity$ is the fluid viscosity, and $\Pressure \equiv \Pressure \SpaceTime$ is the pressure. We assume an isotropic permeability tensor field $\PermTensor \left( \bmSpace \right) = \PermSymbol \left( \bmSpace \right) \bm{I}$, where $\PermSymbol \left( \bmSpace \right)$ is a scalar function and $\bm{I}$ is the identity tensor.

We assume an incompressible fluid flow and use the Boussinesq approximation: density variations are neglected in the mass conservation equation but they are accounted for in the buoyancy term. Consequently, mass conservation equation results in a divergence-free velocity field
\begin{equation}
	\Div \DarcyVeloc = 0.
	\label{app_eq:incompressibilityDim}
\end{equation}

We describe the transport of the fluids using the advection-diffusion equation in terms of the concentration $\Concentration $ of the buoyant fluid. We scale the concentration by the solubility so that $\Concentration \in \left[ 0,1 \right]$. Thus, the transport equation in a Eulerian framework in dimensionless form reads
\begin{equation}
	\Porosity \frac{\partial \Concentration}{\partial \Time} + \DarcyVeloc \cdot \Grad \Concentration - \Porosity \Div \left( \Dispersion \Grad \Concentration \right) = 0,
	\label{app_eq:transportDim}
\end{equation}
where $\Porosity$ denotes the constant medium porosity, and $\Dispersion \equiv \Dispersion \left( \bmSpace \right)$ denotes the dispersion tensor accounting for molecular diffusion and velocity-dependent hydrodynamic dispersion, which represents the spreading due to local flow velocity below the scale of the resolved transport equation \citep{Bear1989}. We assume scalar and constant molecular and hydrodynamic dispersion coefficients. Consequently, the dispersion tensor reduces to $\Dispersion \left( \bmSpace \right) = \Diffusion \bm{I}$, where $\Diffusion$ is the dispersion coefficient.

We scale the spatial variables, viscosity and permeability into dimensionless form as $\bmSpace = \LengthScale \bmSpaceAdim$, $\Viscosity = \ViscosityScale \ViscosityAdim$, and $\PermTensor = \PermScale \PermTensorAdim$, respectively. We denote the spatial operators in dimensionless form with a prime, $\Grad = \LengthScale \GradAdim$ and $\Laplacian = \LengthScale^2 \LaplacianAdim$. Moreover, we consider the following transformations for the density
\begin{equation}
\Density = \DensityScale \DensityAdim + \DensityRef,
\label{eq:densityScaling}
\end{equation}
where $\DensityScale$ is a characteristic density difference (chosen equal to $\DeltaDensitym$ for the unstable case) and $\DensityRef$ equals the density of the denser fluid.
To simplify the buoyancy term, it is useful to transform and scale the pressure as
\begin{equation}
\Pressure = \DensityScale \GravityCt \LengthScale \PressureAdim + \PressureRef,
\label{eq:pressureScaling}
\end{equation}
with the reference pressure being $\PressureRef \left(\bmSpace \right) = - \DensityRef \GravityCt x \sin \Angle - \DensityRef \GravityCt z \cos \Angle $.
Finally, the velocity is scaled with
\begin{equation}
\DarcyVeloc = \DarcyVelocScale \DarcyVelocAdim.
\label{eq:velocityScaling}
\end{equation}
Replacing transformations from \autoref{eq:densityScaling}--\autoref{eq:velocityScaling} in \autoref{app_eq:DarcyDim}, we obtain
\[
\DarcyVelocScale \DarcyVelocAdim = -\frac{\PermScale \DensityScale \GravityCt}{\ViscosityScale} \frac{\PermTensorAdim}{\ViscosityAdim} \left( \GradAdim \PressureAdim - \DensityAdim \GravityUnit \right).
\]
Therefore, choosing the velocity scale as
\[
\DarcyVelocScale = \frac{\PermScale \DensityScale \GravityCt}{\ViscosityScale},
\]
the Darcy's equation in dimensionless form reads
\begin{equation}
	\DarcyVelocAdim = - \frac{\PermTensorAdim}{\ViscosityAdim} \left( \GradAdim \PressureAdim - \DensityAdim \GravityUnit \right).
	\label{app_eq:DarcyAdim}
\end{equation}
Similarly, the incompressibility equation \autoref{app_eq:incompressibilityDim} in dimensionless form reads
\begin{equation}
	\DivAdim \DarcyVelocAdim = 0.
	\label{app_eq:incompressibilityAdim}
\end{equation}

For the transport equation, we scale time as $\Time = \TimeScale \TimeAdim$, where
\[
\TimeScale = \frac{\Porosity \LengthScale}{\DarcyVelocScale}.
\]
Replacing in \autoref{app_eq:transportDim} we obtain
\[
\frac{\partial \Concentration}{\partial \TimeAdim} + \DarcyVelocAdim \cdot \GradAdim \Concentration - \frac{\Diffusion \TimeScale}{\LengthScale^2} \LaplacianAdim \Concentration = 0.
\]
Finally, defining the Rayleigh number as
\begin{equation}
	\Rayleigh = \frac{\LengthScale^2}{\TimeScale \Diffusion} = \frac{\PermScale \DensityScale \GravityCt \LengthScale}{\Porosity \Diffusion \ViscosityScale},
	\label{app_eq:Rayleigh}
\end{equation}
the transport equation in dimensionless form reads
\begin{equation}
	\frac{\partial \Concentration}{\partial \TimeAdim} + \DarcyVelocAdim \cdot \GradAdim \Concentration - \frac{1}{\Rayleigh} \LaplacianAdim \Concentration = 0.
	\label{app_eq:transportAdim}
\end{equation}

\section{Discussion on numerical aspects}
\label{sec:numericsDetails}

\subsection{Mesh convergence study}
\label{sec:numericsDetails_meshConvergence}

In this section, we detail the mesh-convergence study. We consider three meshes with different element sizes: the one used for all the cases reported in the main text with element size $\ElemSize = 0.00125$, a coarser one with element size $2\ElemSize$, and a finer one with element size $\frac{3}{2} \ElemSize$. 

To assess convergence, we compare the root mean square error of the buoyant mass, scalar dissipation rate, interface length, and interface width for different test cases throughout the simulated time interval. We have not considered neither the dissolution rate nor the dissolution flux because they are computed postprocessing other variables. To ensure that the tests are representative homogeneous and heterogeneous unstable cases, we have considered $\Rayleigh=10000$ and, for the heterogeneous case, a permeability field with isotropic correlation structure $\CorrLenx = \CorrLenz = 0.05$ and $\Variance = 1$. We note that the same permeability field has been used for the three meshes — we generate a permeability field for the coarsest mesh and all the finer cells contained inside a coarsest cell have the same value of the permeability as the one assigned to the coarsest cell. Taking as a reference value the one obtained with the finest mesh, for all the tests the mesh of size $\ElemSize$ computes the buoyant mass with an error $\lesssim  10^{-3}$, the scalar dissipation rate $\lesssim 10^{-1}$, the interface length $\lesssim 10^{-3}$, and the interface width $\leq 10^{-1}$ (see \autoref{tab:rmse}).

\begin{table}[h]
    \centering
    \begin{tabular}{lcc|cc}
    & \multicolumn{2}{c|}{\textbf{Homogeneous}} & \multicolumn{2}{c}{\textbf{Heterogeneous}}\\
        \multicolumn{1}{c}{Metric} & Mesh size $2\ElemSize$ & Mesh size $\ElemSize$ & Mesh size $2\ElemSize$ & Mesh size $\ElemSize$\\
        \hline
        Buoyant mass & $3.83 \cdot 10^{-3}$ & $6.04 \cdot 10^{-4}$ & $2.60\cdot 10^{-3}$ & $1.01\cdot 10^{-3}$ \\
        Scalar diss. rate & $1.73 \cdot 10^{-1}$ & $4.76 \cdot 10^{-2}$ & $1.93\cdot 10^{-2}$ & $6.53 \cdot 10^{-3}$ \\
        Interface length & $4.06 \cdot 10^{-3}$ & $8.15 \cdot 10^{-4}$ & $6.62 \cdot 10^{-3}$ & $1.54 \cdot 10^{-3}$ \\
        Interface width & $5.00 \cdot 10^{-2}$ & $3.97 \cdot 10^{-2}$ & $3.62 \cdot 10^{-2}$ & $3.51\cdot 10^{-2}$ \\
    \end{tabular}
    \caption{Relative root mean square error of the buoyant mass, scalar dissipation rate, interface length and interface width for the meshes with element size $2\ElemSize$ and $\ElemSize$, taking as a reference value the one obtained with a mesh of element size $\frac{3}{2}\ElemSize$. Results obtained from a simulation using the unstable density law and $\Rayleigh = 10000$.}
    \label{tab:rmse}
\end{table}

\begin{figure}[h]
        \centering
        \includegraphics[width=\linewidth]{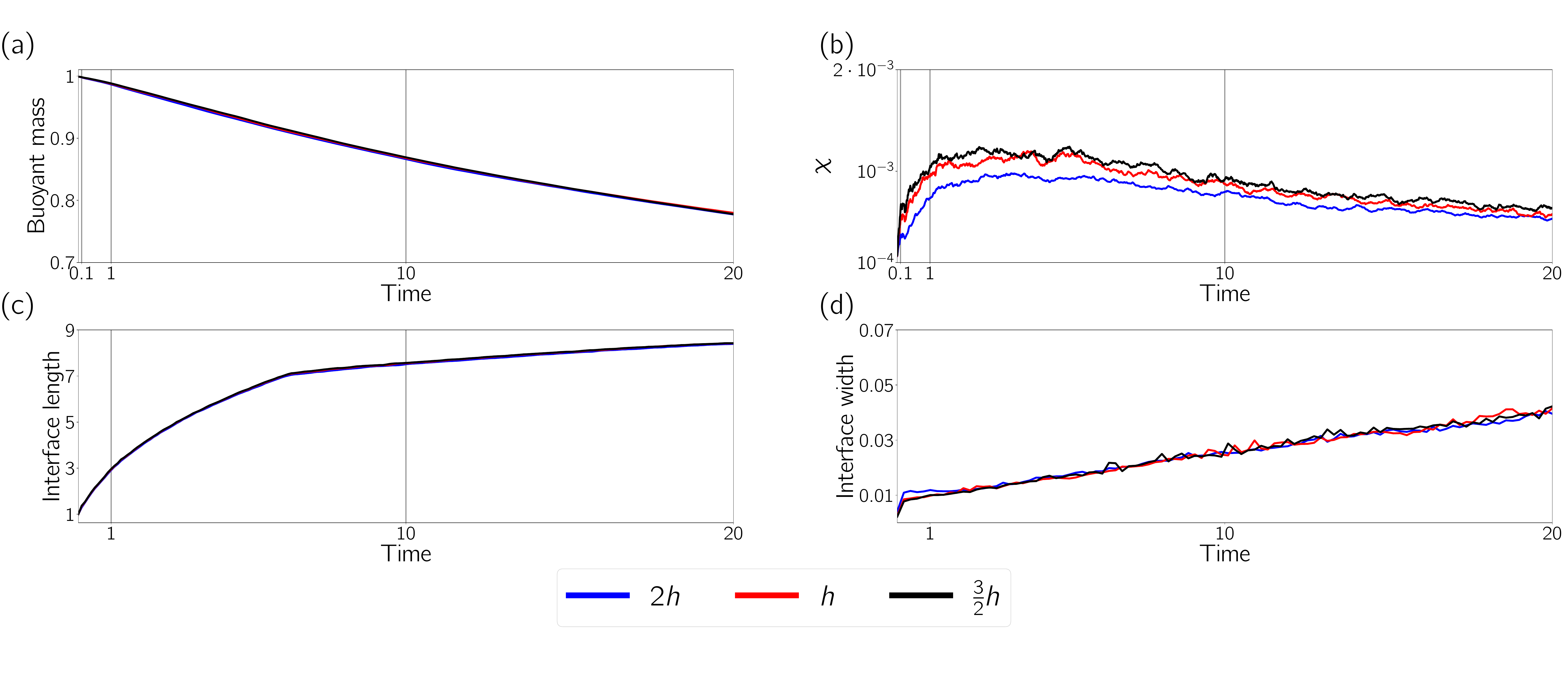}
        {\phantomsubcaption\label{fig:buoyantMass_meshConvergencehetero}}
        {\phantomsubcaption\label{fig:scalarDissipationRate_meshConvergencehetero}}
        {\phantomsubcaption\label{fig:interfaceLength_meshConvergencehetero}}
        {\phantomsubcaption\label{fig:interfaceWidth_meshConvergencehetero}}
    \caption{Time evolution of the \subref{fig:buoyantMass_meshConvergencehetero} buoyant mass, \subref{fig:scalarDissipationRate_meshConvergencehetero} scalar dissipation rate, \subref{fig:interfaceLength_meshConvergencehetero} interface length, and \subref{fig:interfaceWidth_meshConvergencehetero} interface width for a mesh of element size $2\ElemSize$ (blue), $\ElemSize$ (red), and $\frac{3}{2}\ElemSize$ (black) using the unstable density law with $\Rayleigh = 10000$ in a heterogeneous medium with $\CorrLenx = \CorrLenz = 0.05$ and $\Variance = 1$.}
    \label{fig:meshConvergencehetero}
\end{figure}

In \autoref{fig:meshConvergencehetero}, we plot the time evolution of these quantities for the heterogeneous test case. We conclude that the mesh of element size $\ElemSize$ used in this work is sufficient to qualitatively capture the shape of the curves obtained with the more computationally demanding discretization.

\renewcommand{\figsize}{0.32}

\subsection{Initial condition of the concentration}
\label{sec:numericsDetails_initialCondition}

As regards the initial condition for the concentration, to ease numerical convergence of the flux computation at the cell faces and avoid under and overshoots in the concentration value, we smooth the initial interface using a differentiable approximation to the Heaviside step function. The multiplicative factor 100 is a good practical compromise between a fair approximation of the sharp interface and a sufficiently smooth discretization of the transition region ($\approx 20$ cells) avoiding numerical instabilities.
%
%
\paragraph{\textbf{Acknowledgments}}
AJR and JJH acknowledge the support of the MICIU/AEI/10.13039/501100011033 and the European Union NextGenerationEU/PRTR through grant CNS2023-144134 (ESFERA).
%
%
\bibliographystyle{elsarticle-harv}
\bibliography{PorousMedia}
%
%
\end{document}